\documentclass[onecollarge]{svjour2}       % onecolumn "king-size"
\smartqed  % flush right qed marks, e.g. at end of proof
\usepackage[dvipdfmx]{graphicx}
\usepackage{amsmath,amssymb}
\usepackage{mathrsfs}
\usepackage{dcolumn}
\usepackage{bm}
\usepackage{tabularx}
\usepackage{multirow}
\usepackage{color}
\usepackage{ascmac}

\newcommand{\mc}{\mathcal}
\newcommand{\tl}{\tilde}
\newcommand{\ra}{\rangle}
\newcommand{\la}{\langle}
\newcommand{\ve}{\varepsilon}
\newcommand{\eps}{\epsilon}
\newcommand{\sgn}{{\rm \>sgn}}
\newcommand{\tb}{\textbf}
\newcommand{\hxi}{\hat{\xi}}

\newcommand{\CM}{\mathrm{cm}}

\newcommand{\mrG}{\mathrm{G}}

\newcommand{\erfi}{\mathrm{erfi}}
\newcommand{\erf}{\mathrm{erf}}
\newcommand{\nbar}{n_{\rm b}} 
 
\newcommand{\mcG}{\mathcal{G}} 

\newcommand{\hx}{\hat{x}}

\newcommand{\hz}{\hat{z}}

\newcommand{\hr}{\hat{r}}
\newcommand{\hN}{\hat{N}}

\newcommand{\hL}{\hat{L}}

\newcommand{\hb}{\hat{b}}
\newcommand{\ha}{\hat{a}}
\newcommand{\hv}{\hat{v}}
\newcommand{\htau}{\hat{\tau}}
\newcommand{\hT}{\hat{T}}
\newcommand{\heta}{\hat{\eta}}

\newcommand{\mrss}{\mathrm{ss}}
\newcommand{\mrA}{\mathrm{A}}
\newcommand{\sigmaCM}{\sigma_{\CM}}

\def\XXint#1#2#3{{\setbox0=\hbox{$#1{#2#3}{\int}$}
\vcenter{\hbox{$#2#3$}}\kern-.5\wd0}}

\journalname{Journal of Statistical Physics}
\begin{document}

\title{Exact solution to two-body financial dealer model: revisited from the viewpoint of kinetic theory}
\author{Kiyoshi Kanazawa$^{1,2}$ \and Hideki Takayasu$^{3,4}$ \and \\ Misako Takayasu$^4$}
\institute{
			\email{kiyoshi@sk.tsukuba.ac.jp}\\
			$^1$ Faculty of Engineering, Information and Systems, University of Tsukuba, Tennodai, Tsukuba, Ibaraki 305-8573, Japan\\
			$^2$ JST, PRESTO, 4-1-8 Honcho, Kawaguchi, Saitama 332-0012, Japan\\
			$^3$ Sony Computer Science Laboratories, 3-14-13 Higashi-Gotanda, Shinagawa-ku, Tokyo 141-0022, Japan\\
			$^4$ Institute of Innovative Research, Tokyo Institute of Technology, 4259 Nagatsuta-cho, Midori-ku, Yokohama 226-8502, Japan
}

\date{Received: date / Accepted: date}

\maketitle

\begin{abstract}
	The two-body stochastic dealer model is revisited to provide an exact solution to the average order-book profile using the kinetic approach. The dealer model is a microscopic financial model where individual traders make decisions on limit-order prices stochastically and then reach agreements on transactions. In the literature, this model was solved for several cases: an exact solution for two-body traders $N=2$ and a mean-field solution for many traders $N\gg 1$. Remarkably, while kinetic theory plays a significant role in the mean-field analysis for $N\gg 1$, its role is still elusive for the case of $N=2$. In this paper, we revisit the two-body dealer model $N=2$ to clarify the utility of the kinetic theory. We first derive the exact master-Liouville equations for the two-body dealer model by several methods. We next illustrate the physical picture of the master-Liouville equation from the viewpoint of the probability currents. The master-Liouville equations are then solved exactly to derive the order-book profile and the average transaction interval. Furthermore, we introduce a generalised two-body dealer model by incorporating interaction between traders via the market midprice and exactly solve the model within the kinetic framework. We finally confirm our exact solution by numerical simulations. This work provides a systematic mathematical basis for the econophysics model by developing better mathematical intuition. 
\end{abstract}

\section{Introduction}
	Statistical physics is a powerful tool to understand the macroscopic behaviour of physical systems from their microscopic dynamics~\cite{KuboB}, and one of the historical landmarks is the kinetic theory for the Brownian motions~\cite{Chapman1970}. For example, let us consider the dynamics of a small particle in water. This particle experimentally exhibits random motions due to molecular collisions, and such a physical picture is mathematically summarised within the kinetic theory. Indeed, the Langevin equation can be derived from the microscopic physical dynamics via the Liouville equation, Bogoliubov-Born-Green-Kirkwood-Yvon (BBGKY) hierarchy~\cite{Hansen}, Boltzmann equation, and the diffusive limit~\cite{vanKampenB}. 

	Remarkably, the Brownian motions are ubiquitously observed in broad areas, such as in financial markets~\cite{GardinerB}. For example, the price timeseries of stock and foreign exchange rate exhibits random motions similar to the Brownian motion, which was historically pointed out by Bachelier~\cite{Bachelier1900} earlier than Einstein's kinetic theory~\cite{Einstein1905}. Physicists curious about this similarity have studied the financial-market microstructure in terms of statistical physics, hoping to understand its macroscopic behaviour from its microscopic dynamics as a research activity of econophysics~\cite{StanleyB,SlaninaB,BouchaudB}. While there have been various econophysics models proposed, in this report, we focus on a microscopic financial-market model called the dealer model~\cite{Takayasu1992,Sato1998,YamadaPRE2009,KanazawaPRL2018,KanazawaPRE2018}, which depicts the decision-making processes dynamics on the level of individual traders.

	The dealer model is one of the earliest microscopic models that describe individual traders' decision-making process in econophysics~\cite{Takayasu1992}. It was first introduced as a deterministic dynamical model~\cite{Takayasu1992,Sato1998} and was later extended for a stochastic model~\cite{YamadaPRE2009,KanazawaPRL2018,KanazawaPRE2018} for mathematical simplicity. The stochastic dealer model can be mathematically solved for several cases: the two-body case (see Ref.~\cite{YamadaPRE2009} for $N=2$) and the mean-field case (see Refs.~\cite{KanazawaPRL2018,KanazawaPRE2018} for $N\gg 1$) with the total number of traders $N$. The two-body case $N=2$ exactly was solved in Ref.~\cite{YamadaPRE2009}, where the dynamics are finally mapped into the first-hitting time problem to obtain the exact transaction-interval statistics. The mean-field case $N\gg 1$ was solved in Refs.~\cite{KanazawaPRL2018,KanazawaPRE2018} by using the kinetic theory: i.e., the financial Boltzmann and Langevin equations are derived by starting from the financial Liouville equation (i.e., the master equation for the many-body joint probability density function (PDF)) and the BBGKY hierarchy. This method finally deduces the average order-book profile and transaction-interval statistics within the mean-field approximation, which exhibits excellent numerical agreements. In this sense, the dealer model is well-tractable in terms of statistical-physics theories. 

	However, the relationship between the previous two-body solution and the mean-field solution is still elusive since their methodologies are formally different. While the formal BBGKY hierarchy for the two-body case was briefly derived in the supplementary material of Ref.~\cite{KanazawaPRL2018} in a rather incomplete form, its solution was not thoroughly analyzed in Refs.~\cite{KanazawaPRL2018,KanazawaPRE2018} to understand the mathematical characters specific to the two-body case $N=2$, because the main focus of Refs.~\cite{KanazawaPRL2018,KanazawaPRE2018} was to analyze the mean-field case $N\gg 1$. 

	In this report, we revisit the two-body stochastic dealer model to clarify the technical and mathematical roles of the kinetic framework therein. We first derive two exact master-Liouville (ML) equations (i.e., one is a reduced form and the other is the complete form) for the two-body dealer model by fully incorporating the ``collision" mechanism. In particular, we derive the reduced form of the ML equation via two different methods: one is based on Novikov's theorem for the coloured noise, and another is based on the continuous limit from a lattice model. Before the technical derivation, we provide a technical review of Novikov's theorem~\cite{Novikov1965,Hanggi1978} and the Liouville equation for collisional dynamics because they are advanced topics for non-Markovian stochastic processes and kinetic theory\footnote{Experts on these topics can skip this review section because the main-results sections are self-contained. However, we believe such an introductory review section will be helpful for readers unfamiliar with mathematical technicalities since our kinetic theory for financial Brownian motion is very interdisciplinary, requiring the background of econophysics, non-Markovian stochastic processes, and kinetic theory.}. We also illustrate the physical meaning of the ML equations from the viewpoint of the probability current. The probability-current picture will help readers develop better intuition and catch the sense of the specific mathematical forms of the ML equations. Furthermore, the ML equation is exactly solved to obtain the average order-book profile and the average transaction interval. We also examine the consistency between the full and the reduced ML equations. To demonstrate the power of our kinetic framework, we generalise the dealer model to incorporate market-midprice interaction, and solve the model exactly. Finally, we have numerically confirmed the exact solution.

	This report is organised as follows. We explain the mathematical model of the stochastic dealer model for $N=2$ in Sec.~\ref{sec:Model}. In Sec.~\ref{sec:review:technicality}, we provide a technical review on Novikov's theorem and a manipulation technique of the $\delta$-functions for collisional dynamics. In Sec.~\ref{sec:app:der_master_Brownian_confined}, we derive a reduced ML equation exactly by two different approaches. We exactly solve the reduced ML equation in Sec.~\ref{sec:exact_sol_reducedML} to obtain the average order-book profile and the average transaction interval. In Sec.~\ref{sec:fullML_der}, we derive the full ML equation and examine its physical meaning from the viewpoint of the probability current. In addition, we examine the consistency between the reduced and full ML equations. In Sec.~\ref{sec:dealerModel_interaction}, we consider a generalised dealer model by incorporating interaction between traders via the market midprice and exactly solve the generalised model in the kinetic framework again. Finally, we numerically show the numerical simulations to confirm the validity of the exact solutions in Sec.~\ref{sec:numerical_confirmation}. This report is concluded in Sec.~\ref{sec:conclusion} with some remarks. Three appendixes complement these sections. 

\section{Model}\label{sec:Model}
	We formulate the stochastic dealer model based on the Markovian stochastic processes after explaining our mathematical notation.

	\subsection{Mathematical notation}
		Here we briefly explain our mathematical notation. In this report, any stochastic variable accompanies the hat symbol, such as $\hat{A}$ instead of $A$. In addition, the ensemble average of any stochastic variable is denoted by $\la \hat{A}(t)\ra$ at the continuous physical time $t$. All the models are based on the continuous-time Markovian dynamics.  Using this notation and the $\delta$ function\footnote{Readers unfamiliar with the $\delta$ functions are referred to Appendix~\ref{sec:app:delta_func_review}}, the PDF of $\hat{A}(t)$ is given by $P_t(A):=\la\delta(A-\hat{A}(t))\ra$ by stressing the difference between the stochastic variable $\hat{A}(t)$ and the non-stochastic real number $A$. Inversely, the ensemble average $\la\hat{A}(t)\ra$ can be rewritten as
		\begin{equation}
			\la \hat{A}(t)\ra=\int AP_t(A)dA.
			\label{def:ensemble_avg_point}
		\end{equation}

		We use the square brackets for any functional argument, such as $f[\{x(s)\}_{s\in \bm{R}}]$, to stress that $f$ is a functional but not a function, where $\bm{R}:=(-\infty,\infty)$ is the set of the real numbers and $s\in \bm{R}$ is a time point. We sometimes abbreviate the argument, such that $f[x]:=f[\{x(s)\}_s]$ if its meaning is obvious from the context. For any functional, the functional derivative is written by $(\delta f[x])/(\delta x(s))$. 

		In this report, we regard the derivative symbols $\partial/\partial x$ and $\delta/\delta x(t)$ as linear operators acting on all subsequent terms. For example, the derivative $\partial/\partial x$ in the following formula is interpreted as 
		\begin{equation}
			\frac{\partial}{\partial x}\alpha(x)\frac{\partial}{\partial x}\beta (x) P_t(x):= 
			\frac{\partial}{\partial x}\left[\alpha(x)\left\{\frac{\partial}{\partial x}\left(\beta (x) P_t(x)\right)\right\}\right].
		\end{equation}
	
	\subsection{Model dynamics}
		\begin{figure}
			\centering
			\includegraphics[width=150mm]{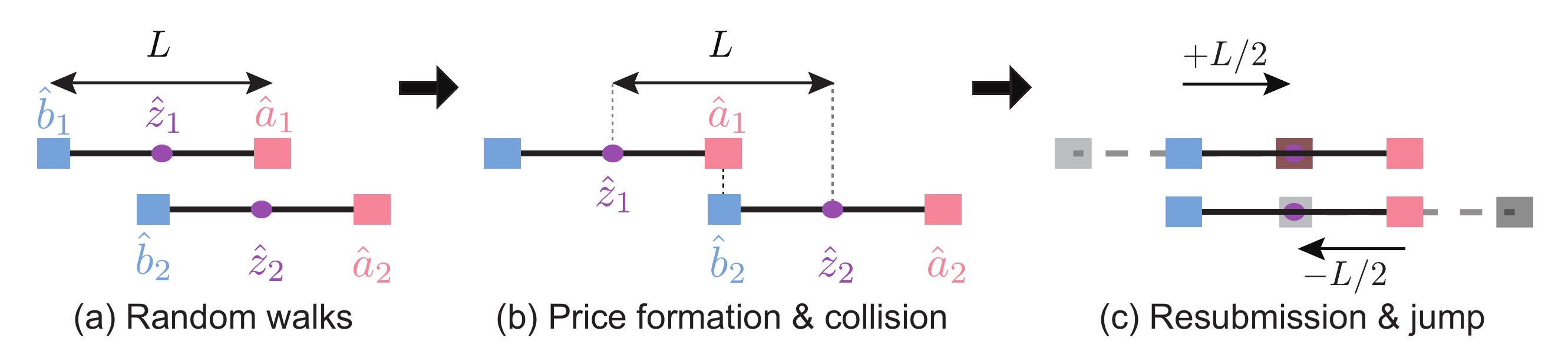}
			\caption{
				Schematic dynamics of the two-body stochastic dealer model. (a)~In the absence of transactions $|\hz_1-\hz_2|<L$, both $\hz_1$ and $\hz_2$ exhibits random walks. (b)~When the condition $|\hz_1-\hz_2|=L$ is met, a transaction happens and thus, the transacted price is updated (i.e., the price formation). These dynamics are essentially similar to the collisions in kinetic theory. (c)~Just after the transaction, both traders resubmit their prices far from the transacted price. This resubmission process is mathematically implemented as jumps of $\hz_1$ and $\hz_2$. 
			}
			\label{fig:2BodyDealerModel}
		\end{figure}
		We consider a market composed of two traders, always quoting both bid and ask prices simultaneously. In this report, $\hb_i(t)$ and $\ha_i(t)$ denotes the bid and ask prices of the $i$th trader for $i=1,2$ at the continuous physical time $t\in \bm{R}$. The difference between the ask and bid prices $\hat{L}_i:=\ha_i-\hb_i>0$ is called the spread of the $i$th trader. We assume that the traders' spreads are the same constant, such that $\hL_i = L = \mbox{const.}>0$ for $i=1,2$ (see Fig.~\ref{fig:2BodyDealerModel}a). We also define the midprices of the trader $i$ as $\hz_i:= (\ha_i+\hb_i)/2$. The transaction condition (see Fig.~\ref{fig:2BodyDealerModel}b) is given by 
		\begin{equation}
			a_1 = b_2 \>\mbox{ or }\> a_2 = b_1 \>\>\> \Longleftrightarrow \>\>\>
			|z_1-z_2| = L. 
		\end{equation}

		In this report, we analyze the simplest case based on the model 1 in Ref.~\cite{YamadaPRE2009} and its generalisation in Sec.~\ref{sec:dealerModel_interaction}. The dynamics of the midprices $\{\hz_i\}_{i=1,2}$ obey the simple Brownian motion in the absence of transactions $|\hz_1-\hz_2|<L$ (see Fig.~\ref{fig:2BodyDealerModel}a):
		\begin{subequations}
			\label{eq:dealermodel}
			\begin{equation}
				\frac{d\hz_i}{dt} = \sigma \hxi^{\mrG}_i
			\end{equation}
			with the white Gaussian noise $\hxi^{\mrG}_i(t)$ and a positive constant $\sigma>0$. The white Gaussian noise is formally defined as the derivative of the Wiener process $\hat{W}_i(t)$ as $\hxi^{\mrG}_i(t):= d\hat{W}_i(t)/dt$ and satisfies
			\begin{equation}
				\la \hxi^{\mrG}_i(t)\ra = 0, \>\>\> \la \hxi^{\mrG}_i(t_1)\hxi^{\mrG}_j(t_2)\ra = \delta(t_1-t_2). 
			\end{equation}
			We note that the higher-order cumulants are absent due to the Gaussian nature\footnote{
				Formally, the $n$th cumulant of any stochastic variable $\hat{A}(t)$ is defined by (see Ref.~\cite{KuboB} for example)
				\begin{equation}
					\la \hat{A}(t_1) \dots \hat{A}(t_n)\ra_{\rm c} := \frac{\delta^n}{\delta \zeta(t_1)\dots \delta \zeta(t_n)}\log \Psi[\zeta], \>\>\> 
					\Psi[\zeta]:= \left<\exp\left[i\int_0^t ds\zeta(s)\hat{A}\right]\right>.
				\end{equation}
				For any Gaussian noise, the higher-order cumulants are absent: $\la \hat{A}(t_1) \dots \hat{A}(t_n)\ra_{\rm c}=0$ for $n\geq 3$.
			}. 
			In the presence of transactions, we assume that both traders requote their prices far from the transaction price to avoid immediate transactions, according to the following equation (see Fig.~\ref{fig:2BodyDealerModel}b and c)
			\begin{equation}
				a_i(t)=b_j(t) \>\>\> \Longleftrightarrow \>\>\> 
				a_i(t+dt) = a_i(t) + \frac{L}{2}, \>\>\> 
				b_j(t+dt) = b_j(t) - \frac{L}{2}
			\end{equation}
			for an infinitesimal positive $dt>0$, or equivalently 
			\begin{equation}
				|z_1(t)-z_2(t)| = L \>\>\> \Longleftrightarrow \>\>\> 
				z_1(t+dt) = z_2(t+dt) = z_1(t) - \frac{L}{2}\sgn(z_1(t)-z_2(t))
			\end{equation}
		\end{subequations}
		with the sign function (i.e., $\sgn(x) = +1$ for $x>0$, $\sgn(x) = 0$ for $x=0$, and $\sgn(x) = -1$ for $x<0$). Remarkably, this transaction rule does not change the centre of mass.
		
		We note that this model is the most basic model; we can introduce additional elements, such as trend-following strategies (see Refs.~\cite{YamadaPRE2009,KanazawaPRL2018,KanazawaPRE2018} for examples). However, since our motivation is to reveal the mathematical characters of the dealer model from the kinetic viewpoint, we start the model as simple as possible and then consider its generalisation in Sec.~\ref{sec:dealerModel_interaction}.

	\subsection{Review of the previous solution based on the first-hitting time problem in Ref.~\cite{YamadaPRE2009}}
		\begin{figure}
			\centering
			\includegraphics[width=150mm]{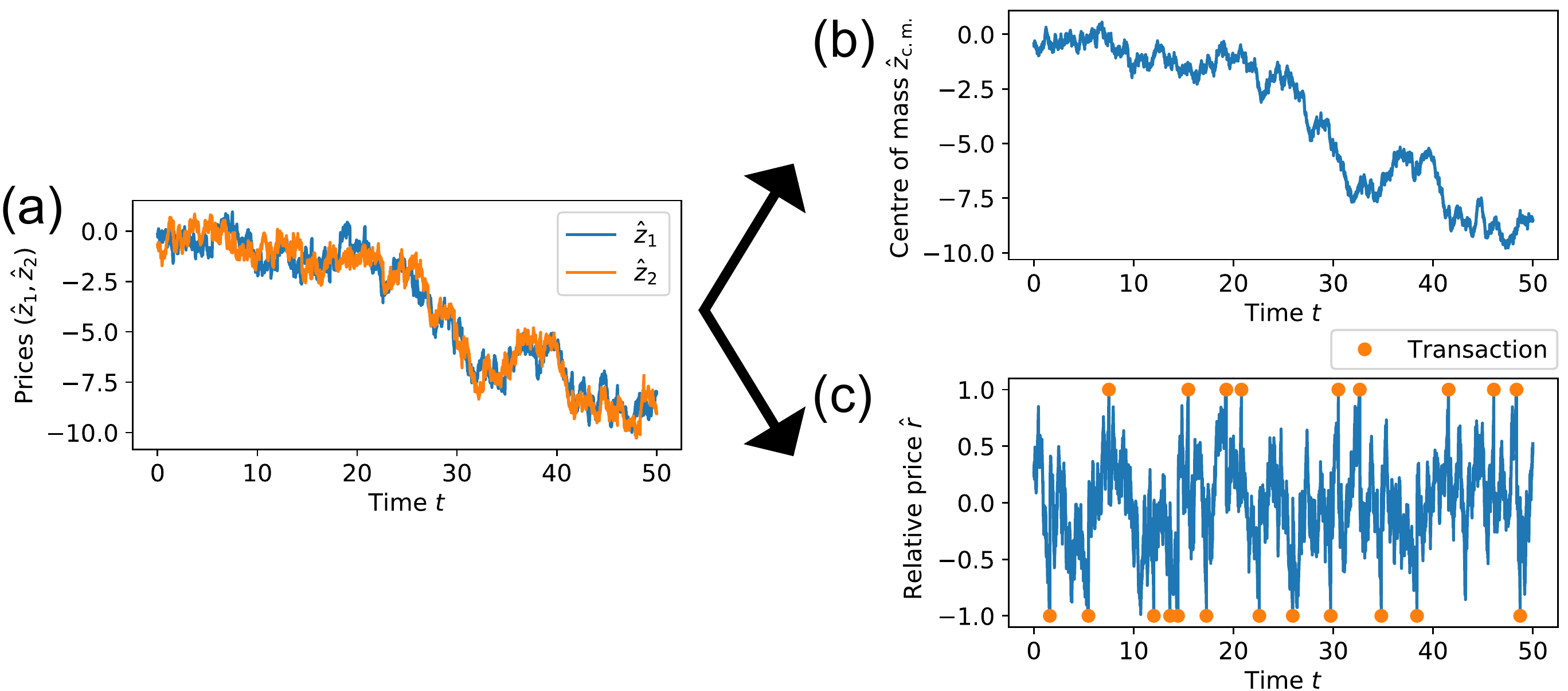}
			\caption{
				Useful representation based on the centre of mass $\hz_{\CM}$ and the relative price $\hr$. (a)~The original dynamics based on $(\hz_1,\hz_2)$ is equivalent to those based on $(\hz_{\CM},\hr)$. (b)~The centre of mass $\hz_{\CM}$ exhibits the simple random walks~\eqref{eq:zCM_varTrans}, irrelevantly to transactions (Fig.~b). (c)~The relative price $\hr$ shows random walks confined in the regime $\hr \in (-L/2,L/2)$. At the boundaries $r=\pm L/2$, the relative price goes back to the origin as described by Eq.~\eqref{eq:r_varTrans} due to transactions. Here we assume $L=2$. 
			}
			\label{fig:varTrans_CM+r}
		\end{figure}
		Here we review the previous solution to this model based on Ref.~\cite{YamadaPRE2009} to clarify the difference to our approach. In Ref.~\cite{YamadaPRE2009}, this model was mapped into the first-hitting time problem of the one-dimensional Brownian motion. Specifically, by introducing the centre of mass $\hz_{\CM}$, the relative price $\hr$ (see Fig.~\ref{fig:varTrans_CM+r}), and the noise variance for the centre of mass $\sigmaCM$, defined by
		\begin{equation}
			\hz_{\CM}:= \frac{\hz_1+\hz_2}{2}, \>\>\>
			\hr:=\hz_1-\hz_{\CM}=\frac{\hz_1-\hz_2}{2}, \>\>\>
			\sigmaCM := \frac{\sigma}{\sqrt{2}},
		\end{equation}
		we obtain
		\begin{subequations}
			\label{eq:dealermodel_trans}
			\begin{screen}
				\begin{align}
					\hz_{\CM}(t+dt) &= \hz_{\CM}(t) + dt\sigmaCM \hat{\chi}(t), 
					\label{eq:zCM_varTrans}\\
					\hr(t+dt) &= 
					\begin{cases}
						\displaystyle \hr (t) + dt\sigmaCM \heta(t) & (|\hr(t)|<L/2) \\
						0 & (|\hr(t)|=L/2)
					\end{cases}
					\label{eq:r_varTrans}
				\end{align}
			\end{screen}
			for an infinitesimal positive timestep $dt>0$. Here the two white noises are introduced as 
			\begin{equation}
				\hat{\chi}(t) := \frac{1}{\sqrt{2}} \left(\hxi_1(t)+\hxi_2	(t)\right), \>\>\> 
				\heta(t) := \frac{1}{\sqrt{2}} \left(\hxi_1(t)-\hxi_2(t)\right),
			\end{equation}	
			which satisfy 
			\begin{equation}
				\la \hat{\chi}(t) \ra = 0, \>\>\> \la \heta(t) \ra = 0
			\end{equation}
			\begin{equation}
				\la \hat{\chi}(t_1)\hat{\chi}(t_2)\ra = \delta(t_1-t_2), \>\>\> 
				\la \heta(t_1)\heta(t_2)\ra = \delta(t_1-t_2), \>\>\> 
				\la \hat{\chi}(t_1)\heta(t_2)\ra = 0. 
			\end{equation}
		\end{subequations}
		We thus find that 
		\begin{enumerate}
			\item the centre of mass $\hz_{\CM}$ obeys the random walk irrelevantly to the transactions because the centre of mass does not move during transactions; 
			\item the relative price $\hr$ obeys the random walk and returns back to the origin if it hits the boundaries at $\hr=\pm L/2$. 
		\end{enumerate}	
		These facts imply that the dynamics of the two-body dealer model is equivalent to the first-hitting time problem at the boundaries $\hr=\pm L/2$. Based on these characters, the authors of Ref.~\cite{YamadaPRE2009} solved this first-hitting time problem to obtain the transaction-interval statistics. For example, the mean transaction interval is given by
		\begin{equation}
			\la \hat{\tau}\ra = \frac{L^2}{2\sigma^2}
			\label{eq:review_YamadaPRE2009_<tau>}
		\end{equation}
		with the transaction interval $\hat{\tau}$. Based on such analytic formulas on the transaction-interval statistics, the authors of Ref.~\cite{YamadaPRE2009} derived the statistical properties of various versions of the dealer model. 

		In addition, the diffusion constant of the centre of mass is exactly given by
		\begin{equation}
			D(N=2):= \frac{\sigmaCM^2(N=2)}{2}, \>\>\> \sigmaCM^2(N=2):= \frac{\sigma^2}{2}. 
		\end{equation}
		This means that the diffusive speed of the centre of mass is slower than that of a single trader. According to the mean-field theory~\cite{KanazawaPRE2018}, in general, the diffusion constant of the centre of mass is given by the scaling 
		\begin{equation}
			D(N) := \frac{\sigmaCM^2(N)}{2}, \>\>\> \sigmaCM^2(N)\propto \frac{\sigma^2}{N},
		\end{equation}
		suggesting that the centre of mass is ``heavier" for large $N$. In this sense, $\sigmaCM(N=2)$ can be regarded as the renormalized diffusion constants as the result of the many-body interaction.

	\subsection{Goal of this report: reformulation based on kinetic theory}
		While the previous derivation is powerful in understanding the two-body dealer model, its relationship is not clear to the mean-field solution proposed in Refs.~\cite{KanazawaPRL2018,KanazawaPRE2018}. The mean-field solution for $N\gg 1$ in Refs.~\cite{KanazawaPRL2018,KanazawaPRE2018} is based on kinetic theory: the microscopic dynamics described by the high-dimensional stochastic differential equations are mapped onto the financial Liouville equation (i.e., an ML equation for the joint PDF), which is reduced to the Boltzmann and Langevin equations via the BBGKY hierarchy. This methodology is very different from the previous first-passage-problem approach, at least superficially. 
		
		This report aims to fill this technical gap mathematically: we derive the Liouville equation for the two-body dealer model exactlyB. Furthermore, the ML equation is solved exactly to obtain the average order-book profile and the mean transaction interval. In addition, we examine the mathematical consistency between these different approaches to develop our better intuition.  

\section{Technical review}\label{sec:review:technicality}
	Here we review Novikov's theorem~\cite{Novikov1965,Hanggi1978} and the collision problem together with related manipulation technique of the $\delta$ functions. These technical methods are relevant to the derivation of the ML equations.

	This section aims to provide a technical background for the general audience unfamiliar with non-Markovian stochastic processes and kinetic theories as an elementary review. Since the main contents are self-contained without needing to refer to this section, professional readers can skip this section and proceed with the main results from Sec.~\ref{sec:app:der_master_Brownian_confined}. However, we believe that such a preliminary review will be helpful for the general audience because the kinetic framework for the dealer model is based on multidisciplinary mathematical backgrounds.

	\subsection{Stochastic calculus for coloured noise: Novikov's theorem}
		\subsubsection{Setup}
			We first review Novikov's theorem~\cite{Novikov1965,Hanggi1978}, a useful theorem for stochastic calculus for coloured noise. This technique will be used for the stochastic dealer model in Sec.~\ref{sec:derMaster-reduced1_Novikov}, and we provide its elementary introduction for a simple case here.
			
			\paragraph{Ornstein-Uhlenbeck coloured noise.}
			Let us first consider an Ornstein-Uhlenbeck (OU) process with a non-zero correlation time $\eps>0$:
			\begin{equation}
				\frac{d\heta_\eps}{dt} = -\frac{\heta_\eps}{\eps} + \hxi^{\mrG}
				\label{sec:Novikov:OU}
			\end{equation}
			with the standard white Gaussian noise $\hxi^{\mrG}$, satisfying $\la\hxi^{\mrG}(t)\ra = 0$ and $\la\hxi^{\mrG}(t_1)\hxi^{\mrG}(t_2)\ra = \delta(t_1-t_2)$. This OU process can be regarded as a specific implementation of the {\it coloured noise} because it satisfies
			\begin{equation}
				\la \heta_{\eps}(t)\ra = 0, \>\>\> \la \heta_{\eps}(t_1)\heta_{\eps}(t_2)\ra = \frac{1}{2\eps}e^{-|t_1-t_2|/\eps},
			\end{equation}
			showing the exponential decay of the correlation with the short characteristic timescale $\eps$. We therefore call $\heta_{\eps}$ the {\it OU coloured noise} and we use the OU coloured noise for a technical calculation of kinetic theory. We note that the OU coloured noise converges to the white noise for the small $\eps$ limit, such that 
			\begin{equation}
				\lim_{\eps\downarrow 0}\la \heta_{\eps}(t_1)\heta_{\eps}(t_2)\ra = \delta(t_1-t_2). 
			\end{equation}

			In addition, the OU coloured noise belongs to the Gaussian noise, which can be proved as follows. It is known that any superposition of Gaussian random variables also obeys the Gaussian statistics~\cite{KuboB}. Because the OU coloured noise can be represented as the linear superposition of the white Gaussian noise, 
			\begin{equation}
				\heta_{\eps}(t) = \heta_{\eps}(0) + \int_0^t e^{-(t-s)/\eps}\hxi^{\mrG}(s)ds,
			\end{equation}
			the OU coloured noise also belongs to the Gaussian noise. 

			\paragraph{Technical advantage of the OU coloured noise.}
			One of the technical advantages to introduce the coloured noise is to remove the singularity of the white noise. Indeed, the white noise is an unbounded and discontinuous function of time, which causes various delicate mathematical issues, such as the It\^o-Stratonovich dilemma~\cite{vanKampen1981}. On the other hand, the OU coloured noise is a bounded and continuous function of time, and various delicate issues disappear for nonzero $\eps>0$. Therefore, in our formulation, we first consider the coloured noise to avoid delicate issues, proceed with calculations for nonzero $\eps>0$, and, finally, take the white-noise limit $\eps\downarrow 0$. 

			\paragraph{Stochastic dynamics under coloured noise.}
			Next, let us consider the stochastic dynamics of a stochastic variable $\hx(t)$ driven by the OU coloured noise:
			\begin{equation}
				\frac{d\hx}{dt} = -\alpha(\hx) + \beta(\hx)\heta_{\eps}
				\label{sec:Novikov:OU_Langevin}
			\end{equation}
			with smooth functions $\alpha(\hx)$ and $\beta(\hx)$. Assuming that the state space is one-dimensional, this system can be regarded as non-Markovian since the OU coloured noise has a finite correlation time. In this sense, the standard mathematical method for the Markovian stochastic processes is not directly available for analytical calculations. 

		\subsubsection{Prescription 1: Markov embedding}
			Because many natural phenomena obey non-Markovian dynamics, various technical methods have been developed in statistical physics for such non-Markovian systems. One established method is based on the Markov embedding technique. This technique is based on converting the original non-Markovian onto an auxiliary higher-dimensional Markovian system by sufficiently adding many auxiliary variables. For example, let us reconsider the non-Markovian dynamics~\eqref{sec:Novikov:OU_Langevin} from the viewpoint of the Markov embedding. While the dynamics~\eqref{sec:Novikov:OU_Langevin} is non-Markovian in the one-dimensional space $\hx(t)$, it can be interpreted as a two-dimensional Markovian dynamics specified by the two-dimensional state vector $\hat{\Gamma}(t):= (\hx(t), \heta_\eps(t))$. Indeed, $\hat{\Gamma}(t)$ obeys the two-dimensional simultaneous SDEs~\eqref{sec:Novikov:OU} and \eqref{sec:Novikov:OU_Langevin} that are Markovian, and thus we can derive the Fokker-Planck equation for the two-dimensional joint PDF $P_t(x, \eta_\eps)$. In this sense, some non-Markovian processes can be converted onto high-dimensional Markovian dynamics by the Markov embedding (e.g., see also Refs.~\cite{ZwanzigB,Kupferman2004,KanazawaPRR2020} for the generalised Langevin equations and Refs.~\cite{KanazawaPRR2020,KanazawaPRL2020,KanazawaNLHawkes2021} for the Hawkes processes). 

		\subsubsection{Prescription 2: functional calculus and Novikov's theorem}
			One of the other famous methods is based the functional calculus for non-Markovian stochastic paths, to which Novikov's theorem~\cite{Novikov1965} belongs. Novikov's theorem is available for coloured Gaussian noises to evaluate ensemble averages (see Ref.~\cite{Hanggi1978} for its generalisation for non-Gaussian coloured noises). This theorem states that an identity
			\begin{subequations}
				\label{eq:NovikovTheorem}
			\begin{screen}
			\begin{equation}
				\la \heta_{\eps}(t)g[\heta_{\eps};t]\ra = \int_{0}^t ds \la \heta_{\eps}(t)\heta_{\eps}(s)\ra \left<\frac{\delta g[\heta_{\eps};t]}{\delta\heta_{\eps}(s)}\right>
				\label{eq:NovikovTheorem1}
			\end{equation}
			\end{screen}
			holds for any functional $g[\heta_{\eps};t]:=g[\{\heta_{\eps}(s)\}_{0\leq s\leq t}]$ dependent on the path of the coloured Gaussian noise $\{\heta_{\eps}(s)\}_{0\leq s\leq t}$. Particularly, for the special case where $g$ is a function of $\hx(t)$, we obtain
			\begin{equation}
				\la \heta_{\eps}(t)g(\hx(t))\ra = \int_{0}^t ds \la \heta_{\eps}(t)\heta_{\eps}(s)\ra \left<\frac{\delta \hx(t)}{\delta\heta_{\eps}(s)}\frac{\partial g(\hx(t))}{\partial \hx}\right>.
				\label{eq:NovikovTheorem2}
			\end{equation}
			\end{subequations}
			Readers unfamilar with the functional derivative $\delta/\delta \heta_{\eps}(s)$ are referred to Appendix.~\ref{sec:app:func_Taylor}. 

			\paragraph{Derivation.}
			This formula can be derived as follows: let us apply the functional Taylor expansion (see Appendix.~\ref{sec:app:func_Taylor}),
			\begin{equation}
				g[\heta_{\eps};t] = g[0;t] + \sum_{n=1}^\infty \frac{1}{n!}\int_0^t d\bm{s}_n
				\frac{\delta^n g[\eta_{\eps};t]}{\delta \eta_{\eps}(s_1)\dots \eta_{\eps}(s_n)}\bigg|_{\eta_{\eps}=0} \heta_{\eps}(s_1)\dots \heta_{\eps}(s_n) , \>\>\>
				d\bm{s}_n := \prod_{k=1}^n ds_k.
				\label{eq:functionalTaylor_1}
			\end{equation}
			By introducing $\bm{s}_n:=(s_1,\dots,s_n)$, this implies 
			\begin{subequations}
			\begin{align}
				\la \heta_{\eps}(t)g[\heta_{\eps};t]\ra &= \sum_{n=1}^\infty \frac{1}{n!} 
				\int_0^t d\bm{s}_n R^{(n)}(\bm{s}_n) 
				\la \heta_{\eps}(t)\heta_{\eps}(s_1)\dots \heta_{\eps}(s_n)\ra, \\
				R^{(n)}(\bm{s}_n) &:= \frac{\delta^n g[\eta_{\eps};t]}{\delta \eta_{\eps}(s_1)\dots \eta_{\eps}(s_n)}\bigg|_{\eta_{\eps}=0}.
			\end{align}
			\end{subequations}
			Here we use a mathematical fact on the Gaussian random numbers:
			\begin{equation}
				\la \heta_{\eps}(t)\heta_{\eps}(s_1)\dots \heta_{\eps}(s_{n})\ra 
				= \sum_{i=1}^{n} \la \heta_{\eps}(t)\heta_{\eps}(s_i)\ra 
				\left< \prod_{j=1 | j\neq i}^n \heta_{\eps}(s_j)\right>.
			\end{equation}
			By considering a symmetry of $R^{(n)}(\bm{s}_n)$
			\begin{equation}
				R^{(n)}(\bm{s}_n) = R^{(n)}(\bm{s}'_n), \>\>\> 
				 \mbox{ where } \bm{s}'_n \mbox{ is any permutation of }\bm{s}_n,
				 \label{eq:symmetry_NovikovTrans}
			\end{equation}
			we obtain 
			\begin{align}
				\la \heta_{\eps}(t)g[\heta_{\eps};t]\ra &= \sum_{n=1}^\infty \frac{1}{n!} 
				\int_0^t d\bm{s}_n R^{(n)}(\bm{s}_n) 
				\sum_{i=1}^{n} \la \heta_{\eps}(t)\heta_{\eps}(s_i)\ra 
				\left< \prod_{j=1 | j\neq i}^n \heta_{\eps}(s_j)\right> \notag \\
				&=  \sum_{n=1}^\infty \frac{1}{n!}\sum_{i=1}^n\int_0^t ds_i \la \heta_{\eps}(t)\heta_{\eps}(s_i)\ra
				\int_0^t \left(\prod_{k=1|k\neq i}^n ds_k\right) 
				R^{(n)}(\bm{s}_n) 
				\left< \prod_{j=1 | j\neq i}^n \heta_{\eps}(s_j)\right> \notag \\
				&= \int_0^t ds_1 \la \heta_{\eps}(t)\heta_{\eps}(s_1)\ra
				\sum_{n=1}^\infty \frac{1}{(n-1)!} \int_0^t ds_2\dots ds_n
				R^{(n)}(\bm{s}_n) 
				\left< \heta_{\eps}(s_2)\dots \heta_{\eps}(s_n)\right>.
				\label{eq:Novikov_trans2}
			\end{align}
			Eq.~\eqref{eq:functionalTaylor_1} implies 
			\begin{align}
				\frac{\delta g[\heta_{\eps};t]}{\delta \heta_{\eps}(s)}
				&= \sum_{n=1}^\infty \frac{1}{n!}\int_0^t d\bm{s}_n
				R^{(n)}(\bm{s}_n) \frac{\delta}{\delta \heta_{\eps}(s)}\{ \heta_{\eps}(s_1)\dots \heta_{\eps}(s_n) \} \notag \\
				&= \sum_{n=1}^\infty \frac{1}{(n-1)!}\int_0^t ds_2\dots ds_n
				R^{(n)}(s,s_2,\dots,s_n) \heta_{\eps}(s_2)\dots \heta_{\eps}(s_n),
				\label{eq:Novikov_trans3}
			\end{align}
			where we have used the symmetry~\eqref{eq:symmetry_NovikovTrans} again. We thus obtain Eq.~\eqref{eq:NovikovTheorem1} by substituing Eq.~\eqref{eq:Novikov_trans3} into Eq.~\eqref{eq:Novikov_trans2}. 

		\subsubsection{Fokker-Planck equation for the white-noise limit} 
			Novikov's theorem is helpful to analyse non-Markovian stochastic processes, in particular for the white-noise limit $\eps\downarrow 0$. Assuming the SDE driven by the OU coloured noise~\eqref{sec:Novikov:OU_Langevin}, let us derive the Fokker-Planck equation for the white-noise limit~\cite{Hanggi1978}. We finally derive the Stratonovich-type Fokker-Planck equation
			\begin{equation}
				\frac{\partial}{\partial t}P_t(x) = \left[\frac{\partial}{\partial x}\alpha(x) + \frac{1}{2}\frac{\partial}{\partial x}\beta(x)\frac{\partial}{\partial x}\beta(x)\right]P_t(x)
				\label{eq:Novikov-StratonovichFP}
			\end{equation}
			for the white-noise limit $\eps\downarrow 0$. 
			
			\paragraph{Derivation.}
			Let us derive the Fokker-Planck equation corresponding to the SDE~\eqref{sec:Novikov:OU_Langevin} in the white-noise limit. The derivative of an arbitary function $f(\hx(t))$ is given by the chain rule\footnote{The ordinary chain rule is available because the OU coloured noise is a continuous and bounded function of time. If we take the white-noise limit first, the ordinary chain rule must be replaced with the It\^o formula.}: 
			\begin{equation}
				\frac{df(\hx)}{dt} = \left(-\alpha(\hx)+\beta(\hx)\heta_{\eps}\right)\frac{df(\hx)}{d\hx}.
			\end{equation}
			Next, let us take the ensemble average of both sides. We first obtain two identities:
			\begin{equation}
				\left<\frac{df(\hx)}{dt}\right> = \lim_{dt\downarrow 0}\left<\frac{f(\hx(t+dt))-f(\hx(t))}{dt}\right> = \int dx f(x)\lim_{dt\downarrow 0}\frac{P_{t+dt}(x)-P_t(x)}{dt} = \int dx f(x)\frac{\partial }{\partial t}P_t(x), 
			\end{equation}
			and 
			\begin{equation}
				\left< \alpha(\hx)\frac{df(\hx)}{d\hx} \right> = \int dxP_t(x)\alpha(x)\frac{df(x)}{dx} 
				= - \int dxf(x)\alpha(x)\frac{\partial P_t(x)}{\partial x},
			\end{equation}
			where we have used the partial integration by assuming the sufficiently-rapid decay of the PDF: 
			\begin{equation}
				\lim_{x\to \pm \infty} P_t(x) = 0, \>\>\> \lim_{x\to \pm \infty} \frac{\partial^n P_t(x)}{\partial x^n} = 0,
			\end{equation}
			with any positive integer $n$. Using Novikov's theorem~\eqref{eq:NovikovTheorem2}, we obtain an identity for the white-noise limit $\eps\downarrow 0$: 
			\begin{align}
				\lim_{\eps\downarrow 0}\left<\heta_{\eps}(t)\beta(\hx(t))\frac{df(\hx(t))}{d\hx}\right> 
				&= \lim_{\eps\downarrow 0}\int_{0}^t ds \frac{1}{2\eps}e^{-(t-s)/\eps} \left<
				\frac{d}{d\hx}\left\{\beta(\hx)\frac{d f(\hx)}{d \hx}\right\}
				\frac{\delta \hx(t)}{\delta\heta_{\eps}(s)} \right> \notag \\
				&= \frac{1}{2} \lim_{s\uparrow t}\left<
				\frac{d}{d\hx}\left\{\beta(\hx)\frac{d f(\hx)}{d \hx}\right\}
				\frac{\delta \hx(t)}{\delta\heta_{\eps}(s)} \right>. 
			\end{align}
			The formal solution of the SDE~\eqref{sec:Novikov:OU_Langevin} is given by
			\begin{equation}
				x(t) = x(t_{\rm ini}) + \int_{t_{\rm ini}}^t dt' \left\{-\alpha(\hx(t'))+\beta(\hx(t'))\heta_{\eps}(t')\right\}
			\end{equation}
			for any $t_{\rm ini} \in [0,t]$. Notably, considering the causality, the effect of $\heta_{\eps}(s)$ is only related to the later value of $\hx(t')$ with $t'>s$. In other words, we obtain 
			\begin{equation}
				\frac{\delta \hx(t')}{\delta \heta_{\eps}(s)} = 0 \>\>\> \mbox{for }t'<s. 
				\label{eq:review:causality_funcDerivative}
			\end{equation}
			Using these relations, we obtain the formal functional derivative 
			\begin{align}
				\frac{\delta \hx(t)}{\delta \heta_{\eps}(s)} &= \int_{t_{\rm ini}}^t dt' \left\{-\frac{\partial \alpha(\hx(t'))}{\partial \hx}\frac{\delta \hx(t')}{\delta \heta_{\eps}(s)}
				+\frac{\partial \beta(\hx(t'))}{\partial \hx}\frac{\delta \hx(t')}{\delta \heta_{\eps}(s)}\heta_{\eps}(t')
				+\beta(\hx(t'))\frac{\delta \heta_{\eps}(t')}{\delta \heta_{\eps}(s)}
				\right\} \notag \\
				&= \int_{t_{\rm ini}}^t dt' \left\{-\frac{\partial \alpha(\hx(t'))}{\partial \hx}
				+\frac{\partial \beta(\hx(t'))}{\partial \hx}\heta_{\eps}(t')\right\}\frac{\delta \hx(t')}{\delta \heta_{\eps}(s)} + 
				\int_{t_{\rm ini}}^t dt' \left\{
				\beta(\hx(t'))\delta(t'-s)
				\right\} \notag \\
				&= \beta(\hx(s)) + \int_{s}^t dt' \left\{-\frac{\partial \alpha(\hx(t'))}{\partial \hx}
				+\frac{\partial \beta(\hx(t'))}{\partial \hx}\heta_{\eps}(t')\right\}\frac{\delta \hx(t')}{\delta \heta_{\eps}(s)}
			\end{align}
			for any $s \in (t_{\rm ini},t)$, where we have used $\delta \heta_{\eps}(t')/\delta \heta_{\eps}(s) = \delta(t'-s)$ and the causality relation~\eqref{eq:review:causality_funcDerivative}. We thus obtain 
			\begin{equation}
				\lim_{s\uparrow t} \frac{\delta \hx(t)}{\delta \heta_{\eps}(s)} = \beta(\hx(t)),
			\end{equation}
			deducing an identity
			\begin{equation}
				\lim_{\eps\downarrow 0}\left<\heta_{\eps}(t)\beta(\hx(t))\frac{df(\hx(t))}{d\hx}\right> 
				= \frac{1}{2} \left<
					\frac{d}{d\hx}\left\{\beta(\hx)\frac{d f(\hx)}{d \hx}\right\}
					\beta(\hx) \right>
				= \frac{1}{2}\int dx f(x)\frac{\partial}{\partial x}\beta(x)\frac{\partial}{\partial x}\beta(x)P_t(x),
			\end{equation}
			where we have used the partial integration implicitly. Since an integral identity
			\begin{equation}
				\int dxf(x)\frac{\partial P_t(x)}{\partial t} = \int dxf(x)
				\left[\frac{\partial}{\partial x}\alpha(x) + \frac{1}{2}\frac{\partial}{\partial x}\beta(x)\frac{\partial}{\partial x}\beta(x)\right]P_t(x)
			\end{equation}
			holds for any function $f(x)$, we obtain the Stratonovich-type Fokker-Planck equation~\eqref{eq:Novikov-StratonovichFP}. 
			
			The resulting Stratonovich-type Fokker-Planck equation~\eqref{eq:Novikov-StratonovichFP} is consistent with the fact that any SDE driven by a coloured Gaussian noise reduces the Stratonovich-type SDE in the white-noise limit~\cite{Wong1965_1,Wong1965_2}. We also note that this result can be confirmed via the Markov embedding approach (i.e., by directly addressing the Fokker-Planck equation for the two-dimensional joint PDF $P_t(x,\eta_{\eps})$) using the projection operators (see Chapter~8 in Ref.~\cite{GardinerB}). 

	\subsection{Collision problem and manipulation technique of $\delta$-functions}
		\label{sec:collision_conventionalKinetic}
		We next review a collision problem and a related technique for the $\delta$ functions, which will be used for the ML equation for the dealer model in Sec.~\ref{sec:app:der_master_Brownian_confined}. Interested readers in the technicality are referred to textbooks on kinetic theory, such as Chapter~13 of Ref.~\cite{BrilliantovB} and Chapter~3 of Ref.~\cite{KanazawaThesis}. 

			\subsubsection{Equation of motion with a deterministic collision}
			\begin{figure}
				\centering
				\includegraphics[width=140mm]{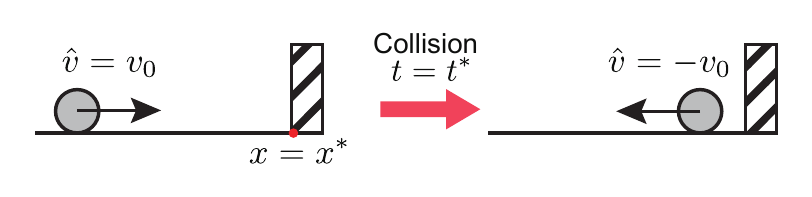}
				\caption{
					Schematic of a single-particle collision. The rigid particle collides against the rigid wall placed at $x=x^*$. Here the particle is regarded as a mass point by ignoring the particle size. After the collision, the particle velocity is inverted as $\hv = -v_0$. The initial condition is given by $(\hx(0),\hv(0))=(x_0,v_0)$ satisfying $x_0<x^*$ and $v_0>0$.
				}
				\label{fig:singleParticleCollision}
			\end{figure}
			This review subsection aims to provide an illustrative example of a deterministic collision based on Ref.~\cite{KanazawaThesis} and then derive the Liouville equation. Let us consider a particle of motion characterised by its position $\hx(t)$ and velocity $\hv(t)$. We place a rigid wall at $x=x^*$ at which the particle's velocity is inverted due to collision (see Fig.~\ref{fig:singleParticleCollision}). The dynamics are given by 
			\begin{subequations}
				\label{eq:collision_eqMotion}
				\begin{align}
					\frac{d\hx}{dt} &= \hv \\
					\frac{d\hv}{dt} &= -2\hv(t)\delta (t-t^*)
				\end{align}				
			\end{subequations}
			with the collision time $t^*$. The collision time $t^*$ is determined by
			\begin{equation}
				\hx(t^*) = x_0 + t^*v_0 = x^*  
				\>\>\>\Longleftrightarrow \>\>\>
				t^* := \frac{x^* - x_0}{v^*}, 
			\end{equation}
			where the initial condition is given by $\hx(0) = x_0$ and $\hv(0) = v_0$ satisfying $x_0 < x^*$ and $v_0 > 0$. Here the multiplication between $\hv(t)$ and the $\delta$ function is interpreted in the It\^o sense: $\hv(t)\delta(t-t^*) = \lim_{h\downarrow 0}\hv(t^*-h)\delta(t-t^*)$. The $\delta$ function represents the impulsive force due to the collision consistently with the invertion. Indeed, this time-evolution equation consistently deduces  
			\begin{equation}
				\hv(t^*+h) = \hv(t^*-h) + \int_{t^*-h}^{t^*+h} dt \frac{d\hv(t)}{dt} = \hv(t^*-h) + \int_{t^*-h}^{t^*+h} dt \left\{-2\hv(t)\delta (t-t^*)\right\} = - \hv(t^*-h)
			\end{equation}
			for infinitesimal positive $h>0$. 

			\subsubsection{Liouville equation}
			From Eq.~\eqref{eq:collision_eqMotion}, we derive the Liouville equation\footnote{Technically, this is called the {\it pseudo-Liouville equation}~\cite{BrilliantovB,KanazawaThesis}. While the conventional Liouville equation is local (i.e., a partial differential equation without jump terms) in classical mechanics, the pseudo-Liouville equation involves non-local jumps due to collisions.}, a time-evolution equation for the two-dimensional joint PDF $P_t(x,v):= \la \delta (x-\hx(t))\delta(v-\hv(t))\ra$:
			\begin{equation}
				\frac{\partial P_t(x,v)}{\partial t} = -v\frac{\partial P_t(x,v)}{\partial x} + 
				\left[\Theta(v)P_t(x,v)-\Theta(-v)P_t(x,-v)\right]|v| \delta(x-x^*). 
				\label{eq:Liouville-simpleCollision}
			\end{equation}
			
			\paragraph{Derivation.}
			For an arbitrary function $f(\hx,\hv)$, its time-evolusion during $[t,t+dt)$ is given by 
			\begin{equation}
				df(\hx,\hv) = 
				\begin{cases}
					\displaystyle 
					\hv(t)\frac{\partial f(\hx(t),\hv(t))}{\partial \hx}dt & (t^*\not \in [t,t+dt)) \\
					f(\hx(t),-\hv(t)) - f(\hx(t),\hv(t)) & (t^*\in [t,t+dt))
				\end{cases}
			\end{equation}
			with $df(\hx,\hv):= f(\hx(t+dt),\hv(t+dt)) - f(\hx(t),\hv(t))$ and infinitesimal positive $dt>0$. This relation is equivalent to 
			\begin{equation}
				\frac{df(\hx,\hv)}{dt} = \hv\frac{\partial f(\hx,\hv)}{\partial \hx} + \left[f(\hx,-\hv) - f(\hx,\hv)\right]\delta(t-t^*).
			\end{equation}
			Let us take the ensemble average of both sides to obtain 
			\begin{equation}
				\left< \frac{df(\hx,\hv)}{dt} \right> = \left< 
					\hv\frac{\partial f(\hx,\hv)}{\partial \hx} + \left[f(\hx,-\hv) - f(\hx,\hv)\right]\delta(t-t^*)
				\right>.
			\end{equation}
			Here we consider a relation, 
			\begin{equation}
				\left< \hv\frac{\partial f(\hx,\hv)}{\partial \hx} \right>
				= \int_{-\infty}^\infty dxdv P_t(x,v)v\frac{\partial f(x,v)}{\partial x}
				= -\int_{-\infty}^\infty dxdv f(x,v)v\frac{\partial P_t(x,v)}{\partial x}
			\end{equation}
			using the partial integration. 
			
			We next make a transformation of the term $\la [f(\hx,\hv)-f(\hx,-\hv)]\delta(t-t^*)\ra$ using a useful idenity
			\begin{equation}
				\delta (\hx(t)-x^*) = \frac{1}{|d\hx(t)/dt|}\delta (t-t^*) = \frac{1}{|\hv(t)|} \delta (t-t^*), \>\>\> \mbox{assuming $\hx(t)<x^*$ and $\hv(t)>0$}.
				\label{eq:collision_direction_conventional_kinetic}
			\end{equation}
			Here the condition $\hv(t)>0$ is essential because it restricts the collision direction.  This implies 
			\begin{equation}
				\delta(t-t^*) = |\hv(t)| \Theta(\hv(t))\delta(\hx(t)-x^*) 
				\>\>\>
				\mbox{with the Heaviside function }\Theta(x):= 
				\begin{cases}
					1 & (x>0) \\ 
					1/2 & (x=0) \\
					0	& (x<0)
				\end{cases}. 
				\label{eq:delta-decompose-simpleKinetic}
			\end{equation}
			We obtain
			\begin{align}
				&\la [f(\hx,\hv)-f(\hx,-\hv)]\delta(t-t^*)\ra \notag \\
				=& \int_{-\infty}^\infty dxdvP_t(x,v)\left[f(x,v)-f(x,-v)\right]|v| \Theta(v)\delta(x-x^*) \notag \\
				=& \int_{-\infty}^\infty dxdvf(x,v)\left[\Theta(v)P_t(x,v)-\Theta(-v)P_t(x,-v)\right]|v| \delta(x-x^*), 
			\end{align}
			where we have made a variable transformation $-v \to v$ for the second term. We thus obtain an identity for any $f(x,v)$:
			\begin{align}
				&\int_{-\infty}^\infty dxdvf(x,v)\frac{\partial P_t(x,v)}{\partial t} \notag \\ 
				= &\int_{-\infty}^\infty dxdvf(x,v)\left\{
					-v\frac{\partial P_t(x,v)}{\partial x} + 
					\left[\Theta(v)P_t(x,v)-\Theta(-v)P_t(x,-v)\right]|v| \delta(x-x^*)
				\right\},
			\end{align}
			leading the Liouville equation~\eqref{eq:Liouville-simpleCollision}. 
			
			It is straightforward to generalise this calculation for multi-body particle dynamics in principle by assuming the rigid spheres without contact friction, finally leading to the Liouville equation for many-body dynamics. In the standard program of the kinetic theory, the Boltzmann equation is finally deduced by integrating the Liouville equation with the assumption of molecular chaos.

		\subsubsection{Technical keys for the derivation: useful identities for the $\delta$ functions}
			The key technique for the derivation is the decomposition of the $\delta$ function~\eqref{eq:delta-decompose-simpleKinetic}. In general, the following identity holds for the $\delta$-functions for an arbitrary function $g(t)$: 
			\begin{screen}
			\begin{equation}
				\delta(g(t)) = \sum_{i}\frac{1}{|g'(t_i)|}\delta(t-t_i)
			\end{equation}
			\end{screen}
			with the $i$th zero point $t_i$, by assuming $g'(t_i)\neq 0$. This technique is straightforwardly applied for the derivation of the ML equation for the dealer model in Sec.~\ref{sec:dynamics-f(r(t))-forMasterEq)}. Readers unfamiliar with the $\delta$ function are referred to Appendix.~\ref{sec:app:delta_func_review}.

\section{Result 1: a reduced master-Liouville equation}
\label{sec:app:der_master_Brownian_confined}
	We have provided enough preliminary review on Novikov's theorem and kinetic theory in Sec.~\ref{sec:review:technicality}. This section derives the master-Liouville equations for the two-body dealer model \eqref{eq:dealermodel} or equivalently \eqref{eq:dealermodel_trans} using kinetic theory as the main results.

	In this section, in particular, we focus on the ML equation for the relative price $\hr$ obeying Eq.~\eqref{eq:r_varTrans}. Finally, we will obtain a reduced ML equation as follows: 
	\begin{screen}
	\begin{subequations}
		\label{eq:masterEq1_confined}
		The time-evolution equation of the reduced PDF $P_t(r)$ is given by  
		\begin{align}
			\frac{\partial P_t(r)}{\partial t} &= \frac{\sigmaCM^2}{2}  \frac{\partial^2}{\partial r^2}P_t(r) + \sum_{s=\pm 1}[J_{t;s}(r+sL/2) - J_{t;s}(r)], \\
			J_{t;s}(r) &:= -s\frac{\sigmaCM^2}{2}\partial_{-s} P_t(sL/2)\delta(r-sL/2) = \frac{\sigmaCM^2}{2}\left|\partial_{-s} P_t(sL/2)\right|\delta(r-sL/2) \geq 0
		\end{align}
		for the one-dimensional PDF $P_t(r):=\la \delta(r-\hr(t))\ra$. The positive term $J_{t;s}(r)$ represents the {\it probability current} due to collisions, as will be discussed in detail in Sec.~\ref{sec:interpretation_probCurrent}. 
	\end{subequations}
	\end{screen}

	Here we have introduced the left ($s=-1$) and right ($s=+1$) derivatives, defined by
	\begin{equation}
		\partial_{s}f(r) := \lim_{h\downarrow 0}\frac{f(r+sh)-f(r)}{sh}. 
	\end{equation}
	Since the original dynamics~\eqref{eq:dealermodel} is two-dimensional specified by the full PDF $P_t(z_1,z_2):=\la \delta(z_1-\hz_1(t))\delta(z_2-\hz_2(t))\ra$, the one-dimensional PDF $P_t(r):=\la \delta(r-\hr(t))\ra$ can be called a {\it reduced PDF}. Correspondingly, we call Eq.~\eqref{eq:masterEq1_confined} a {\it reduced master-Liouville equation} or just a reduced ML equation\footnote{The time-evolution equation of PDFs is regularly called the {\it master equation}, the {\it differential form of the Chapman-Kolmogorov equation} or {\it Kolmogorov's forward equation}. Here we call such an equation the {\it master-Liouville equation} because it corresponds to the Liouville equation in the standard program of kinetic theory.}. 

	Here we provide two independent derivations on the reduced ML equation~\eqref{eq:masterEq1_confined}: one is based on (i)~Novikov's theorem for coloured noise~\cite{Novikov1965,Hanggi1978} and (ii)~continuous limit from one-dimensional random walks on a regular lattice. While all methods deduce the same result, we believe that presenting various derivation methods will help the reader better understand our methodology.

	\subsection{Derivation based on Novikov's theorem for coloured noise}
		\label{sec:derMaster-reduced1_Novikov}
		\begin{figure}
			\centering
			\includegraphics[width=100mm]{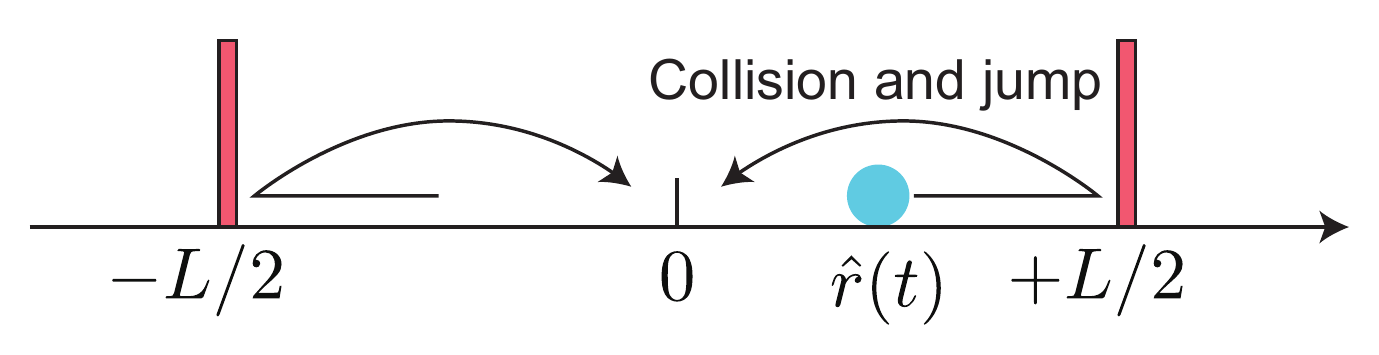}
			\caption{The dynamics of the relative price $\hr(t)$ can be regarded as a mechanical system where the particle diffuses and goes back to the origin after colliding against the boundaries at $r=\pm L/2$.}
			\label{fig:BrownianConfined}
		\end{figure}
		Here we derive the ML equation~\eqref{eq:masterEq1_confined} using Novikov's theorem~\cite{Novikov1965,Hanggi1978} for the SDE~\eqref{eq:r_varTrans} for the relative price $\hr(t)$. Interestingly, the SDE~\eqref{eq:r_varTrans} can be regarded as a Brownian motion confined in the boundaries at $r=\pm L/2$, at which the particle goes back to the origin (see Fig.~\ref{fig:BrownianConfined}). By regarding the particle arrivals at the boudaries $r=\pm L/2$ as ``collisions", this system can be regarded as a mechanical system with collisions and jumps at the boundaries. This is the basic idea to apply the kinetic formulation to this dealer model. 

		This physical picture based on collisional jumps are very similar to the conventional kinetic formulation, and derivation of the ML equation should work well in the parallel method presented in Sec.~\ref{sec:collision_conventionalKinetic}. However, if we naively follow the same formal calculation as that in Sec.~\ref{sec:collision_conventionalKinetic}, we will encounter delicate mathematical issues originating from the $\delta$ singularity of the white noise. To solve this problem via kinetic theory, we reformulate the SDE~\eqref{eq:r_varTrans} by introducing three tricks:
		\begin{enumerate}
			\item replacement of the white noise $\heta(t)$ with the OU coloured noise $\heta_{\eps}(t)$ with nonzero correlation time $\eps>0$;
			\item manipulation of the $\delta$ functions to represent the boundary condition; 
			\item the white-noise limit $\eps\downarrow 0$ to keep the consistency with the original model. 
		\end{enumerate}
		Here the OU coloured noise is introduced to avoid technical delicate issues on the white noise. The OU coloured noise has the advantages that it is a bounded and continuous function of time and that the ordinary calculus is available for formal calculations by assuming non-zero $\eps>0$. In the final stage of the calculation, we take the white-noise limit $\eps\downarrow 0$. Let us explain this procedure one by one as follow. 

		\subsubsection{Reformulation of the SDE for the relative price $\hr(t)$}
			To follow to these recipes, let us first replace the white noise $\heta$ in Eq.~\eqref{eq:r_varTrans} with the OU coloured noise $\heta_{\eps}$ defined by
			\begin{equation}
				\frac{d\heta_\eps(t)}{dt} = -\frac{\heta_{\eps}(t)}{\eps} + \hxi^{\mrG}(t)
			\end{equation}
			with a nonzero positive constant $\eps>0$ and a white Gaussian noise $\hxi^{\mrG}(t)$. This OU noise satisfies 
			\begin{equation}
				\la \heta_{\eps}(t)\ra = 0, \>\>\>
				\la \heta_\eps(t_1)\heta_\eps(t_2)\ra=\frac{1}{2\eps}e^{-|t_1-t_2|/\eps},
			\end{equation}
			which reduces the white noise for $\eps\downarrow 0$:
			\begin{equation}
				\lim_{\eps\downarrow 0}\la \heta_\eps(t_1)\heta_\eps(t_2)\ra = \delta(t_1-t_2). 
			\end{equation}
			We then rewrite Eq.~\eqref{eq:r_varTrans} as 
			\begin{equation}
				d\hr(t) = \begin{cases}
					dt\sigmaCM\heta_{\eps}(t) & (\mbox{without collisions: }\htau_{s;i} \not \in [t,t+dt)) \\
					-sL/2 & (\mbox{with a collision: }\htau_{s;i} \in [t,t+dt))
				\end{cases}, 
			\end{equation}
			with $d\hr(t):=\hr(t+dt)-\hr(t)$ for an infinitesimal positive $dt>0$. 
			Here $\htau_{s;i}$ is the $i$th arrival time of the particle at $\hr(\tau_{s;i})=sL/2$ for $s =\pm 1$. Using the $\delta$ function, this is equivalent to 
			\begin{equation}
				\frac{d\hr(t)}{dt} = \sigmaCM\heta_{\eps}(t) 
				+ \sum_{s=\pm 1}\sum_{i=1} \frac{-sL}{2}\delta(t-\htau_{s;i}). 
				\label{eq:SDE_reduced_r_3}
			\end{equation}
			Considering the collison directions, we note that the velocity must be positive (negative) at the arrival time at $r=+L/2$ ($r=-L/2$): 
			\begin{equation}
				\lim_{h\downarrow 0}\frac{d\hr(\htau_{+1;i}-h)}{dt} = \sigmaCM\lim_{h\downarrow 0}\heta_{\eps}(\htau_{+1;i}-h)> 0, \>\>\> 
				\lim_{h\downarrow 0}\frac{d\hr(\htau_{-1;i}-h)}{dt} = \sigmaCM\lim_{h\downarrow 0}\heta_{\eps}(\htau_{-1;i}-h) < 0.
			\end{equation}
			In other words, 
			\begin{equation}
				s\lim_{h\downarrow 0}\frac{d\hr(\htau_{s;i}-h)}{dt} = s\sigmaCM\lim_{h\downarrow 0}\heta_{\eps}(\htau_{s;i}-h)> 0.
				\label{eq:velocity_direction_at_Collision}
			\end{equation}
			for both $s=\pm 1$. Remarkably, this mathematical structure is essentially similar to the conventional kinetic theory for collision (see Eq.~\eqref{eq:collision_direction_conventional_kinetic} in Sec.~\ref{sec:collision_conventionalKinetic}). This technical issue is important in removing the absolute operators of the OU coloured noises as shown in Sec.~\ref{sec:dynamics-f(r(t))-forMasterEq)}. 
		
		\subsubsection{Dynamics of an arbitrary function $f(\hr(t))$}
			\label{sec:dynamics-f(r(t))-forMasterEq)}
			We also consider the dynamics of an arbitrary function $f(r)$ 
			\begin{equation}
				df(\hr) = \begin{cases}
										\displaystyle 
										\sigmaCM \frac{df(\hr)}{dr}\heta_\eps dt & (\mbox{without collisions: }\htau_{s;i} \not \in [t,t+dt)) \\
										f(\hr-sL/2) -f(\hr) & (\mbox{with a collision: }\htau_{s;i}  \in [t,t+dt))
									\end{cases}
				\label{eq:master_der_trans_cases_111}
			\end{equation}
			with $df(r(t)):=f(r(t+dt))-f(r(t))$. 
			Using the $\delta$-function, this relation can be rewritten as
			\begin{equation}
				\frac{df(\hr)}{dt} 
				= \underbrace{\sigmaCM  \frac{df(\hr)}{dr}\heta_\eps}_{\mbox{without collisions during $[t,t+dt)$}} + \sum_{s=\pm 1}\sum_{i=1}\underbrace{\left[f(\hr-sL/2)-f(\hr)\right]\delta(t-\htau_{s;i})}_{\mbox{with collisions}}. 
			\end{equation}
			Here we use a decomposition formula for the $\delta$ functions:
			\begin{equation}
				\delta(g(t)) = \sum_{i}\frac{1}{|g'(t)|}\delta(t-t_i)
				\>\>\> \Longrightarrow \>\>\>
				h(t)|g'(t)|\delta(g(t_i)) = \sum_{i}h(t)\delta(t-t_i)
			\end{equation}
			for an arbitrary functions $g(t)$ and $h(t)$, where $t_i$ is the $i$th zero point of $g(t)$, defined by $g(t_i)=0$ and $t_i < t_{i+1}$. Using this formula, we obtain 
			\begin{align}
				\frac{df(\hr)}{dt} 
				&= \sigmaCM  \frac{df(\hr)}{dr}\heta_\eps + \sigmaCM\sum_{s=\pm 1}\left[f(\hr-sL/2)-f(\hr)\right] \left|\frac{d\hr}{dt}\right|\delta(\hr-sL/2) \notag \\
				&= \sigmaCM  \frac{df(\hr)}{dr}\heta_\eps + \sigmaCM\sum_{s=\pm 1}\left[f(\hr-sL/2)-f(\hr)\right] |\heta_{\eps}|\delta(\hr-sL/2).
				\label{eq:app:trans_delta_master1}
			\end{align}
			By considering the physical picture that the velocity $d\hr/dt$ must be positive (negative) when the particle hits the boundary $\hr=L/2$ ($\hr=-L/2$) as summarised in Eq.~\eqref{eq:velocity_direction_at_Collision}, we can remove the absolue operator\footnote{
				The introduction of the OU coloured noise $\heta_{\eps}$ with nonzero $\eps>0$ plays a technically important role here. If we assumed the white noise from the beginning, there appeared the absolute values of the white noise $|\hxi^{\mrG}|$, which is mathematically ill-defined. In addition, the collisional velocities~\eqref{eq:velocity_direction_at_Collision} should have diverged for the white noise. To avoid these delicate issues, we proceed with the calculation for nonzero $\eps>0$ and finally take the white-noise $\eps\downarrow 0$. 
			} in Eq.~\eqref{eq:app:trans_delta_master1} such that $|\heta_{\eps}|\delta(\hr-sL/2)=s\heta_{\eps}\delta(\hr-sL/2)$, leading to
			\begin{equation}
				\frac{df(\hr)}{dt} 
				= \sigmaCM  \frac{df(\hr)}{dr}\heta_\eps + \sigmaCM\sum_{s=\pm 1}s\left[f(\hr-sL/2)-f(\hr)\right]\heta_{\eps}\delta(\hr-sL/2).
			\end{equation}
		
		\subsubsection{Ensemble average and Novikov's theorem for the white-noise limit}
			Let us take the ensemble average:
			\begin{equation}
				\left< \frac{df(\hr)}{dt} \right> 
				=\left<\sigmaCM\frac{df(\hr)}{d\hr}\heta_\eps + \sigmaCM\sum_{s=\pm 1}s\left[f(0)-f(sL/2)\right]\heta_{\eps}\delta(\hr-sL/2) \right>.
				\label{eq:master_trans_der_1928}
			\end{equation}
			Let us evaluate the diffusive term $\la (df(\hr)/d\hr)\heta_\eps\ra$ and the collision term $\la \heta_{\eps}\delta(\hr-sL/2)\ra$ one by one. 

			\paragraph{Novikov's theorem for short-time interval.}
				To evaluate the ensemble averages, we use Novikov's theorem that is valid for an arbitrary coloured Gaussian noise~\cite{Novikov1965,Hanggi1978}: 
				\begin{equation}
					\la \heta_{\eps}(t)g(\hr(t)) \ra = \int_0^t dt'\la\heta_{\eps}(t)\heta_{\eps}(t')\ra 
					\left<\frac{\delta g(\hr(t))}{\delta \heta_{\eps}(t')}\right>.
				\end{equation}
				Since the correlation time $\eps$ is finally set infinitesimal, we evaluate this ensemble average by dropping minor correction terms disappearing for the white noise limit $\eps\downarrow 0$. Here we remark a useful identity identity for integrals with short memory: 
				\begin{equation}
					\lim_{\eps\downarrow 0}\int_{0}^t dt'\frac{e^{-(t-t')/\eps}}{2\eps}f(t')
					= \lim_{\eps\downarrow 0}\int_{t_{\rm ini}(\eps)}^t dt'\frac{e^{-(t-t')/\eps}}{2\eps}f(t') = f(t), \>\>\>
					t_{\rm ini}(\eps):= t-\eps^{1/2}
				\end{equation}
				for any function $f(t)$ that decays sufficiently rapidly for $t\to \infty$. This relation holds because
				\begin{equation}
						\lim_{\eps\downarrow 0}\int_{0}^t dt'f(t')
								\frac{e^{-(t-t')/\eps}}{2\eps}
						=  \lim_{\eps\downarrow 0}\int_{0}^{\infty}
								d\tilde{t} f(t-\eps\tilde{t})e^{-\tilde{t}}/2 
						= f(t)
				\end{equation}
				and
				\begin{equation}
						\lim_{\eps\downarrow 0}\int_{t_{\rm ini}(\eps)}^t dt'f(t')
								\frac{e^{-(t-t')/\eps}}{2\eps}
						=  \lim_{\eps\downarrow 0}\int_{0}^{\eps^{-1/2}}
								d\tilde{t} f(t-\eps\tilde{t}) e^{-\tilde{t}}/2 
						= f(t)
				\end{equation}
				with the variable transformation $\tilde{t}:=(t-t')/\eps$. This result is intuitively reasonable because the integral interval $t-t_{\rm ini}(\eps)=\eps^{1/2}$ is much larger than the correlation time $\eps$: $\eps^{1/2}\gg \eps$ for $\eps\downarrow 0$. 
				
				Considering this relationship, it is sufficient to consider the contribution of the path $\{\hr(t'')\}_{t''\in [t_{\rm ini}(\eps),t)}$ with sufficiently short-time interval $t-t_{\rm ini}(\eps)=\eps^{1/2}$:
				\begin{equation}
					\lim_{\eps\downarrow 0}\la \heta_{\eps}(t)g(\hr(t)) \ra = \lim_{\eps\downarrow 0}\int_{t_{\rm ini}(\eps)}^{t} dt'\la\heta_{\eps}(t)\heta_{\eps}(t')\ra 
					\left<\frac{\delta g(\hr(t))}{\delta \heta_{\eps}(t')}\right>.
					\label{eq:Novikov_VshortInterval}
				\end{equation}
				This formula implies that only short-time information on the path $\{\hr(t'')\}_{t''\in [t_{\rm ini}(\eps),t)}$ is necessary in evaluating the ensemble average for the white-noise limit.  

			\paragraph{Rearranged collision timeseries.}
				\begin{figure}
					\centering
					\includegraphics[width=90mm]{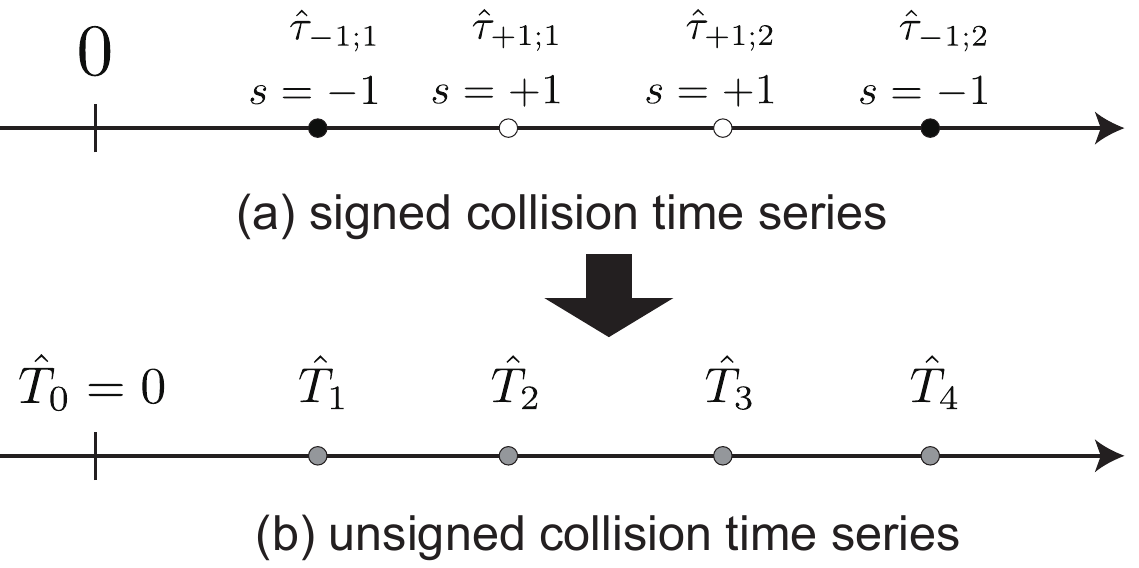}
					\caption{
						The unsigned timeseries $\{\hT_k\}_k$ is introduced as the rearrangement of the signed timeseries $\{\htau_{s;k}\}_{s=\pm 1;k}$. 
					}
					\label{fig:rearrangedTimeSeries}
				\end{figure}
				Here we reformulate the mathematical notation of the collision times. We have already introduced the signed collision timeseries $\{\htau_{s;i}\}_{i}$ for both $s=\pm 1$. To define the unsigned collision timeseries $\{\hT_i\}_i$, let us rearrange these collision timeseries in the ascending order without considering the sign $s=\pm 1$ (see Fig.~\ref{fig:rearrangedTimeSeries}): 
				\begin{equation}
					\mbox{for any $i=1,2,...$, there exists $k$ and $s$, such that }
					\hT_i = \htau_{s;k},
					\mbox{ satisfying }\hT_i < \hT_{i+1}.
				\end{equation}
				We also assume $\hT_0=0$. Using this notation, the presense of collision at time $t$ can be written as the condition $t=\hT_i$ for some $i$, whereas the absense of collision can be written as $t\neq \hT_i$ for any $i$. 

				Notably, if there is a collision at time $t$ (i.e., $t= \hT_i$ for some $i$), the collision term $[f(0) -f(sL/2)]\delta(\hr -sL/2)$ is much more dominant than the diffusive term $\sigmaCM(df/d\hr)\heta_{\eps}$ in Eq.~\eqref{eq:master_trans_der_1928}, such that $|\sigmaCM(df/d\hr)\heta_{\eps}| \ll |[f(0) -f(sL/2)]\delta(\hr -sL/2)|$, due to the $\delta$ singularity. Therefore, at time $t$, we can assume
				\begin{itemize}
					\item the absense of collisions (i.e., $t\neq \hT_i$ for any $i$) in evaluating $\la \sigmaCM(df/d\hr)\heta_{\eps} \ra$, and 
					\item the presense of a collision (i.e., $t= \hT_i$ for some $i$) in evaluating $\la[f(0) -f(sL/2)]\delta(\hr -sL/2)\ra$.
				\end{itemize}
				This means that the dominant term is switched whether there is a collision at the time $t$ or not. 

				We also estimate the probability of a collision during a infinitesimal-time interval $[t,t+dt)$. The probability $p_{\rm col}(t,t+dt)$ should be proportional to $dt$, such that 
				\begin{equation}
					p_{\rm col}(t,t+dt)=\lambda_{\rm col}(t) dt+ o(dt)
				\end{equation}
				with intensity $\lambda_{\rm col}(t)$. Indeed, if $p_{\rm col}(t,t+dt)\propto dt^m$ with $m < 1$, the expected number of collisions $\hN(T)$ during $[0,T)$ is estimated as 
				\begin{equation}
					\la\hN(T) \ra \propto \frac{T}{dt}dt^{m} = Tdt^{m-1}, 
				\end{equation}
				which diverges to infinity for $dt\downarrow 0$. 

			\paragraph{Evaluation of the diffusive term.}
				Based on Novikov's theorem~\eqref{eq:Novikov_VshortInterval} for short-time interval $t-t_{\rm ini}(\eps)$, we evaluate the diffusive term $\la \sigmaCM(df(\hr)/d\hr)\heta_\eps\ra $ as 
				\begin{equation}
					\lim_{\eps\downarrow 0}\left<\sigmaCM\frac{df(\hr(t))}{d\hr}\heta_\eps\right> 
					= \lim_{\eps\downarrow 0} 
					\int_{t_{\rm ini}(\eps)}^t dt'\frac{e^{-(t-t')/\eps}}{2\eps}
					\left<\sigmaCM\frac{\delta \hr(t)}{\delta \heta_{\eps}(t')}\frac{\partial}{\partial \hr(t)}\left(\frac{df(\hr(t))}{d\hr(t)}\right)\right>. 
				\end{equation}
				We then evaluate the functional derivative of the path $\delta \hr(t)/\delta \heta_{\eps}(t')$, by assuming $t\neq \hT_i$ for any $i$. Since the integral interval $t - t_{\rm ini}(\eps)=\eps^{1/2}$ is to be set infinitesimal, we can assume the absense of collisions during $[t_{\rm ini}(\eps),t)$, such that $\hT_{i-1}<t_{\rm ini}(\eps)<t<\hT_{i}$. 
				
				This assumption can be quantitatively discussed by considering the statistical number of collisions during $[t_{\rm ini}(\eps),t)$. Let us introduce $p_k(t_{\rm ini}(\eps),t)$ and $\la \dots \ra_{k;[t_{\rm ini}(\eps),t)}$ as the probability of $k$-times collisions during $[t_{\rm ini}(\eps),t)$ and the ensemble average conditional on $k$-times collisions, respectively. The ensemble average can be written as
				\begin{equation}
					\left<\sigmaCM\frac{\delta \hr(t)}{\delta \heta_{\eps}(t')}\frac{\partial}{\partial \hr(t)}\left(\frac{df(\hr(t))}{d\hr(t)}\right)\right> 
					= \sum_{k=0}^\infty p_k(t_{\rm ini}(\eps),t) \left<\sigmaCM\frac{\delta \hr(t)}{\delta \heta_{\eps}(t')}\frac{\partial}{\partial \hr(t)}\left(\frac{df(\hr(t))}{d\hr(t)}\right)\right>_{k;[t_{\rm ini}(\eps),t)}.
				\end{equation}
				Since the interval is proportional short $t-t_{\rm ini}(\eps)=\eps^{1/2}$, the probability of $k$-times collisions during $[t_{\rm ini}(\eps),t)$ is estimated to be $[\lambda_{\rm col}(t) (t-t_{\rm ini}(\eps))]^k\propto \eps^{k/2}$. For $\eps\downarrow 0$, it is sufficient to consider the case $k=0$ as the leading-order contribution. 				
				
				Therefore, by assuming the absense of collisions (i.e., $k=0$), the path is given by 
				\begin{equation}
					\hr(t) = \hr(t_{\rm ini}(\eps)) + \int_{t_{\rm ini}(\eps)}^t dt' \sigmaCM \heta_{\eps}(t'),
					\label{eq:sol_path_woCollision0}
				\end{equation}
				as the formal solution of the SDE~\eqref{eq:SDE_reduced_r_3}. This implies the functional-derivative relationship
				\begin{equation}
					\frac{\delta \hr(t)}{\delta \heta_{\eps}(t')} = \sigmaCM,
				\end{equation}
				leading to 
				\begin{equation}
					\lim_{\eps\downarrow 0}\left<\sigmaCM\frac{df(\hr(t))}{d\hr}\heta_\eps(t) \right> 
					= \lim_{\eps\downarrow 0} 
					\int_{t_{\rm ini}(\eps)}^t dt'\frac{e^{-(t-t')/\eps}}{2\eps}
					\left<\sigmaCM^2\frac{\partial}{\partial \hr}\left(\frac{df(\hr)}{d\hr}\right)\right>
					= \frac{\sigmaCM^2}{2} \left<\frac{d^2f(\hr)}{d\hr^2}\right>.
				\end{equation}

			\paragraph{Evaluation of the collision term.}
				We next evaluate the collisional term $\la \heta_{\eps}(t)\delta(\hr(t)-sL/2) \ra$, by assuming $t= \hT_i$ for some $i$. Using Novikov's theorem~\eqref{eq:Novikov_VshortInterval} for short-time interval, we obtain 
				\begin{align}
					\lim_{\eps\downarrow 0}\la \heta_{\eps}(t)\delta(\hr(t)-sL/2) \ra 
					= \lim_{\eps\downarrow 0}\int_{t_{\rm ini}(\eps)}^t dt'\frac{e^{-(t-t')/\eps}}{2\eps}
					\left<\frac{\delta \hr(t)}{\delta \heta_{\eps}(t')}\frac{\partial}{\partial \hr(t)}\delta(\hr(t)-sL/2)\right>.
					\label{eq:app:trans_Novikov_jammpingPic_master3}
				\end{align}
				Since the integral interval $t - t_{\rm ini}(\eps)=\eps^{1/2}$ is to be set infinitesimal, we can assume the absense of collisions during $[t_{\rm ini}(\eps),t)$, such that $\hT_{i-1}<t_{\rm ini}(\eps)<t=\hT_{i}$, similarly to the diffusive term. Therefore, the formal solution of the SDE~\eqref{eq:SDE_reduced_r_3} and its functional derivative are given by
				\begin{equation}
					\hr(t) = \hr(t_{\rm ini}(\eps)) + \int_{t_{\rm ini}(\eps)}^{t} dt' \sigmaCM \heta_{\eps}(t'), \>\>\> \Longleftrightarrow \>\>\> 
					\frac{\delta \hr(t)}{\delta \heta_{\eps}(t')} = \sigmaCM.
					\label{eq:sol_path_woCollision}
				\end{equation}
				We then obtain
				\begin{align}
					\lim_{\eps\downarrow 0}\la \heta_{\eps}(t)\delta(\hr(t)-sL/2) \ra 
					&= \lim_{\eps\downarrow 0}\int_{t_{\rm ini}(\eps)}^t dt'\frac{e^{-(t-t')/\eps}}{2\eps}\left<\sigmaCM\frac{\partial}{\partial \hr(t)}\delta(\hr(t)-sL/2)\right> \notag \\
					&= \frac{\sigmaCM}{2}\left<\frac{\partial}{\partial \hr(t)}\delta(\hr(t)-sL/2)\right> \notag \\
					&= -\frac{\sigmaCM}{2}  \frac{\partial}{\partial r}P_t(sL/2).
				\end{align}

				\paragraph{Obtaining the ML equation.}
				In summary, we obtain
				\begin{equation}
					\int_{-\infty}^\infty dr' \frac{\partial}{\partial t}P_t(r')f(r')
					=\int_{-\infty}^{\infty}dr'\left\{P_t(r')\frac{\sigmaCM^2}{2}  \frac{d^2}{dr'^2}f(r')\right\}
					-\frac{\sigmaCM^2}{2}\sum_{s=\pm 1}s\left[f(0)-f(sL/2)\right]\frac{\partial}{\partial r}P_t(sL/2).
				\end{equation}
				By substituting $f(r') = \delta(r'-r)$ and performing the partial integration, we obtain
				\begin{subequations}
				\begin{align}
					\frac{\partial}{\partial t}P_t(r)
					=&\frac{\sigmaCM^2}{2}  \frac{\partial^2}{\partial r^2}P_t(r) - \frac{\sigmaCM^2}{2}\sum_{s=\pm 1}s\left[\delta(r)-\delta(r-sL/2)\right]\frac{\partial}{\partial r}P_t(sL/2) \notag \\
					=&\frac{\sigmaCM^2}{2}  \frac{\partial^2}{\partial r^2}P_t(r) + \sum_{s=\pm 1}\left[J_{t;s}(r+sL/2)-J_{t;s}(r)\right]
					\label{eq:master_reduced_trans1}
				\end{align}
				with the probability current due to collisions 
				\begin{equation}
					J_{t;s}(r):= -s\frac{\sigmaCM^2}{2}\frac{\partial P_t(sL/2)}{\partial r}\delta(r-sL/2).
					\label{eq:master_reduced_trans1_2}
				\end{equation}
				\end{subequations}
				While this is a valid time-evolution equation of the PDF $P_t(r)$, we next rewrite Eq.~\eqref{eq:master_reduced_trans1} to a more intuitve form by examining several technical issues.

		\subsubsection{Technical issue on the left and right derivatives}
			\begin{figure}
				\centering
				\includegraphics[width=80mm]{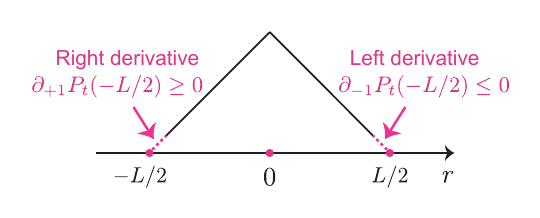}
				\caption{
					Technically, the derivatives $\partial P_t(L/2)/\partial r$ and $\partial P_t(-L/2)/\partial r$ should be regarded as left and right derivatives, respectively. Indeed, considering the obvious character $P_t(r)=0$ for $r>L/2$, $\partial_{+1} P_t(L/2)=0$ and $\partial_{-1} P_t(-L/2)=0$. In addition, the signs of the derivatives are given by $\partial_{-1} P_t(L/2)\leq 0$ and $\partial_{+1} P_t(-L/2)\geq 0$. 
				}
				\label{fig:left_and_right_derivative}
			\end{figure}
			Technically, the derivative $\partial P_t(sL/2)/\partial r$ in the reduced ML equation~\eqref{eq:master_reduced_trans1} should be interpreted as the left (right) derivatives for positive (negative) $s$ (see Fig.~\ref{fig:left_and_right_derivative}), such that
			\begin{equation}
				\frac{\partial P_t(sL/2)}{\partial r} \to \partial_{-s} P_t(sL/2), \>\>\> 
				\partial_{s}f(r) := \lim_{h\downarrow 0}\frac{f(r+sh)-f(r)}{sh}, 
			\end{equation}
			because the probability must be exactly zero beyond the boundary
			\begin{equation}
				P_t(r) = 0 \>\>\> \left(\mbox{for }|r|\geq L/2 \right).
			\end{equation}
			In other words, the particle must come from left (right) when it collides against the right (left) boundary $r=L/2$ ($r=-L/2$), which is reflected for the selection of the derivative direction. 

		\subsubsection{Sign of derivatives}
			We further examine the sign of the derivatives in rewriting Eq.~\eqref{eq:master_reduced_trans1}. Since $P_{t}(r)=0$ for $|r|\geq L/2$, we obtain the signs of the left and right derivative (see Fig.~\ref{fig:left_and_right_derivative}) as 
			\begin{equation}
				\partial_{-1} P_t(+L/2) \leq 0,\>\>\>
				\partial_{+1} P_t(-L/2) \geq 0\>\>\>
				\Longleftrightarrow \>\>\>
				s\partial_{-s} P_t(sL/2) = -|\partial_{-s}P_t(sL/2)|.
			\end{equation}
			Indeed, since $P_t(r)=0$ for $|r|\geq L/2$ and $P_t(r)\geq 0$ for $|r|<L/2$, we can show  
			\begin{subequations}
			\begin{align}
				\partial_{-1}P_t(L/2) &= \lim_{h\downarrow 0}\frac{P_t(L/2-h)-P_t(L/2)}{-h} = -\lim_{h\downarrow 0}\frac{P_t(L/2-h)}{h} \leq 0, \\
				\partial_{+1}P_t(L/2) &= \lim_{h\downarrow 0}\frac{P_t(L/2+h)-P_t(L/2)}{h} = +\lim_{h\downarrow 0}\frac{P_t(L/2+h)}{h} \geq 0.
			\end{align}
			\end{subequations}
			This means that the probability current $J_{t;s}(r)$ defined by Eq.~\eqref{eq:master_reduced_trans1_2} can be rewritten as an explicitly positive form
			\begin{align}
				J_{t;s}(r) := \frac{\sigmaCM^2}{2}|\partial_{-s}P_t(sL/2)|\delta(r-sL/2) \geq 0. 
			\end{align}
			This is equivalent to the reduced ML equation~\eqref{eq:masterEq1_confined}, which is a more intuitive form than Eq.~\eqref{eq:master_reduced_trans1} because the direction of the probability current is clear as discussed in Sec.~\ref{sec:interpretation_probCurrent}.  

	\subsection{Intuitive interpretation of the reduced master-Liouville equation~\eqref{eq:masterEq1_confined}}
			\label{sec:interpretation_probCurrent}
			\begin{figure}
				\centering
				\includegraphics[width=150mm]{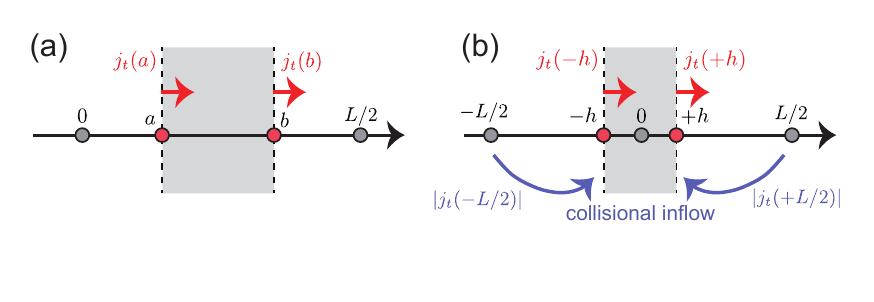}
				\caption{
					Interpretation of the ML equation~\eqref{eq:masterEq1_confined_j_reformulated} based on the probability current $j_t(r)$ defined by Eq.~\eqref{eq:def_probCurrent}. (a)~$j_t(r)$ represents the probability current because the probability conservation relation is written as $(d/dt)\int_{a}^{b} drP_t(r) = -j_t(b) + j_t(a)$ for $a,b\in (0,L/2)$. (b)~The $\delta$ contributions from the term $J_{t;s}(r)$ represents the collisional inflow at $r=\pm L/2$ because the probability conservation relation is written as $(d/dt)\int_{-h}^{h} drP_t(r) = -j_t(h) + j_t(-h) + |j_t(-L/2)| + |j_t(L/2)|$. These schematic illustrates the intuitive meaning of the $\delta$ contributions in $J_{t;s}(r)$. 
				}
				\label{fig:prob_current}
			\end{figure}
			Here we provide an intuitive interpretation of the master-Louville equation~\eqref{eq:masterEq1_confined} from the viewpoint of the probability inflow and outflow relevant to the probability conservation. 
			\begin{screen}
				By introducing the {\it probability current}~\cite{GardinerB} defined by 
				\begin{subequations}
					\label{eq:reducedML_probCurrentForm} 
					\begin{equation}
						j_t(r):= - \frac{\sigmaCM^2}{2}\frac{\partial P_t(r)}{\partial r},
						\label{eq:def_probCurrent} 
					\end{equation}
					the reduced ML equation~\eqref{eq:masterEq1_confined} can be rewritten as 
					\begin{equation}
						\frac{\partial P_t(r)}{\partial t} = -\frac{\partial}{\partial r}j_t(r) + \sum_{s=\pm 1}[J_{t;s}(r+sL/2) - J_{t;s}(r)], \>\>\> 
						J_{t;s}(r) := |j_t(r)|\delta(r-sL/2) \geq 0.
						\label{eq:masterEq1_confined_j_reformulated}
					\end{equation}
				\end{subequations}
			\end{screen}

			Here we have abbreviated the technical minor symbol on the left and right derivatives. The term $j_t(r)$ is called the probability current because the probability-conservation relation 
			\begin{equation}
				\frac{d}{dt}\int_{a}^{b} dr P_t(r) = \int_{a}^b\left[-\frac{\partial}{\partial r}j_t(r)\right] = - \underbrace{j_t(b)}_{\rm outflow} + \underbrace{j_t(a)}_{\rm inflow}
			\end{equation}
			implies that the total probability within the interval $(a,b)$ is determined by the balance of the probability outflow $j_t(b)$ and inflow $j_t(a)$ (see Fig.~\ref{fig:prob_current}a for a schematic), where $(a,b)$ is any interval which does not include the singular points $r=0$ and $r=\pm L/2$, such that $0<a<b<L/2$ or $-L/2<a<b<0$.
			
			Considering the meaning of the probability current $j_t(r)$, the $\delta$-type contribution can be interpreted as follows: let us consider the probability conservation near $r=0$ by integrating the ML equation~\eqref{eq:masterEq1_confined_j_reformulated} over $(-h,h)$ as 
			\begin{align}
				\frac{d}{dt}\int_{-h}^{h} dr P_t(r) = \underbrace{- j_t(h) + j_t(-h)}_{\mbox{diffusive flow}} + \underbrace{|j_t(+L/2)| + |j_t(-L/2)|}_{\mbox{collisional inflow}}
			\end{align}
			with infinitesimal positive $h>0$. This means that the $\delta$-contribution from $J_{t;s}(r)$ is to transfer the collisional probability current at $r=\pm L/2$ to the origin $r=0$ (see Fig.~\ref{fig:prob_current}b for a schematic). This picture illustrates the balance of the probability currents described by the ML equation~\eqref{eq:masterEq1_confined_j_reformulated}, consistently with the physical dynamics where the particle returns back to the origin after collision. 
			
			In addition, let us consider the probability conservation near the wall at $r=L/2$, by integrating the ML equation~\eqref{eq:masterEq1_confined_j_reformulated} over $(L/2-h,L/2+h)$ as 
			\begin{align}
				\frac{d}{dt}\int_{L/2-h}^{L/2+h} dr P_t(r) = - j_t(L/2+h) + j_t(L/2-h) -|j_t(L/2)| = 0 + \underbrace{j_t(L/2-h)}_{\mbox{diffusive inflow}} -\underbrace{j_t(L/2)}_{\mbox{collisional outflow}},
			\end{align}
			where we have used $j_t(L/2+h)=0$ and $j_t(L/2)>0$. We thus find that the total probability within the interval $(L/2-h,L/2+h)$ is determined by the balance between the diffusive inflow $j_t(L/2-h)$ and the collisional outflow $j_t(L/2)$.

	\subsection{Derivation based on the continuous limit from a lattice model}
		We have derived the ML equation~\eqref{eq:masterEq1_confined} via Novikov's theorem, which requires advanced techniques on both non-Markovian stochastic processes and kinetic theory. However, considering that the derivation is multidisciplinary, we believe that another more elementary derivation without requiring such technicalities will be helpful for non-experts. Therefore, we provide a more elementary derivation of the reduced ML equation~\eqref{eq:masterEq1_confined} based on a random-walk model on the one-dimensional lattice. 

		\begin{figure}
			\centering
			\includegraphics[width=120mm]{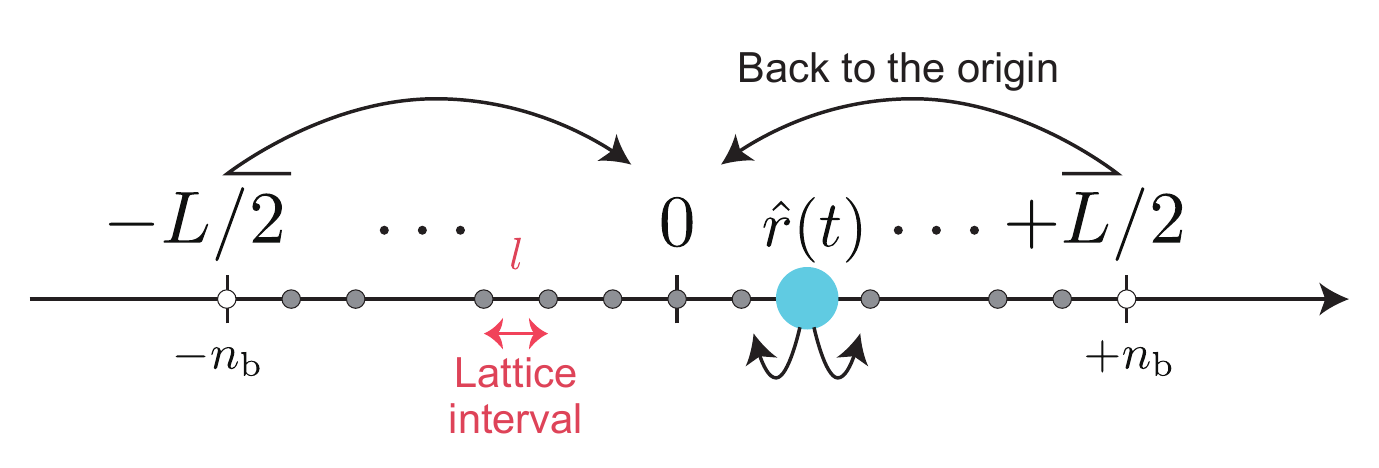}
			\caption{
				Discrete random walks model~\eqref{eq:SDE_discrete} on a lattice with interval $l$. The particle $\hr(t)$ hops to the nearest-neighbors $\hr(t)-l$ and $\hr(t)+l$ with equal probability, according to the Poisson process of intensity $\lambda$. The particle goes back to the origin at the boundary $|\hr(t)|=L/2$ (equivalently, when $|\hr(t)|=l\nbar$, the particle goes back to the origin with the intensity $\lambda/2$). 
			}
			\label{fig:lattice}
		\end{figure}
		Let us consider a lattice of interval $l$ and the particle dynamics on the lattice (see Fig.~\ref{fig:lattice}). The location of the particle is denoted by $\hr(t)$, which obeys 
		the symmetric Poisson jump process whose jump sizes are $\pm l$ in the absense of collisions against barriers, such that
		\begin{equation}
			\hr(t+dt) = \hr(t) + \begin{cases}
				+ l & (\mbox{prob.}=\frac{\lambda}{2} dt) \\
				- l & (\mbox{prob.}=\frac{\lambda}{2} dt) \\
				0 & (\mbox{prob.}=1-\lambda dt)
			\end{cases}
		\end{equation}
		with an infinitesimal positive $dt>0$ and the total Poisson intensity $\lambda>0$. We note that the probability of occurrence of a Poisson jump during $[t,t+dt)$ is given by $\lambda dt$. The hopping barriers are placed at $r=\pm L/2:=\pm l\nbar$ with positive integer $\nbar$. At the moment of collisions $\hr(t)=\pm L/2$, the particle comes back to the origin $r=0$. In summary, the stochastic dynamics is given by 
		\begin{equation}
			\hr(t+dt) = 
			\begin{cases}
				\hr(t) + l & (\mbox{prob.}=\frac{\lambda}{2} dt,\mbox{ if } \hr(t) \leq l(\nbar-2)) \\
				\hr(t) - l & (\mbox{prob.}=\frac{\lambda}{2} dt,\mbox{ if } \hr(t) \geq l(2-\nbar)) \\
				0 & (\mbox{prob.}=\frac{\lambda}{2} dt,\mbox{ if } |\hr(t)| = l(\nbar-1)) \\
				\hr(t) & (\mbox{prob.}=1-\lambda dt)
			\end{cases}. 
			\label{eq:SDE_discrete}
		\end{equation}
		This system is a discrete version of the SDE~\eqref{eq:r_varTrans} and reduces to the SDE~\eqref{eq:r_varTrans} for the continuous limit, as will be discussed later. 
		
		\subsubsection{Master-Liouville equation for the lattice model}
			Let us derive the ML equation for this lattice model as follow. For an arbitrary function $f(\hr(t))$, we obtain the following identity, 
			\begin{align}
				df(\hr(t)) = 
				\begin{cases}
					f(\hr(t) + l)-f(\hr(t)) & (\mbox{prob.}=\frac{\lambda}{2} dt,\mbox{ if } \hr(t) \leq l(\nbar-2)) \\
					f(\hr(t) - l)-f(\hr(t)) & (\mbox{prob.}=\frac{\lambda}{2} dt,\mbox{ if } \hr(t) \geq l(2-\nbar)) \\
					f(0) - f(\hr(t)) & (\mbox{prob.}=\frac{\lambda}{2} dt,\mbox{ if } |\hr(t)| = l(\nbar-1)) \\
					0 & (\mbox{prob.}=1-\lambda dt)
				\end{cases}. 
			\end{align}
			with $df(\hr(t)):=f(\hr(t+dt))-f(\hr(t))$.
			By taking the ensemble average of both sides, we obtain
			\begin{equation}
				\left<\frac{df(\hr)}{dt}\right> = \left< 
					\frac{\lambda}{2}\left[f(\hr+l)-f(\hr)\right]\bm{1}_{D_+}(\hr) + 
					\frac{\lambda}{2}\left[f(\hr-l)-f(\hr)\right]\bm{1}_{D_-}(\hr) + 
					\frac{\lambda}{2}\left[f(0)-f(\hr)\right]\bm{1}_{D_{\partial}}(\hr)
				\right>
				\label{eq:app:lattice_masterEq}
			\end{equation}
			with the indicator function $\bm{1}_D(x)$ for the following domain symbols
			\begin{equation}
				D_+:= (-\infty,l(\nbar-2)], \>\>\>
				D_-:= [-l(\nbar-2), \infty), \>\>\>
				D_{\partial}:= \{r=\pm l(\nbar-1)\}. 
			\end{equation}
			The indicator function $\bm{1}_D(x)$ is introduced for any domain $D$, such that 
			\begin{equation}
				\bm{1}_{D}(x) = \begin{cases}
					1 & (x \in D) \\
					0 & (x \not \in D)
				\end{cases}.
			\end{equation}
			Equation~\eqref{eq:app:lattice_masterEq} is equivalent to
			\begin{align}
				&\sum_{r}f(r)\frac{\partial P_t(r)}{\partial t} \notag \\
				= &\frac{\lambda}{2}\sum_r P_t(r)\left[
					\left\{f(r+l)-f(r)\right\}\bm{1}_{D_+}(r)
					+\left\{f(r-l)-f(r)\right\}\bm{1}_{D_-}(r) 
					+\left\{f(0)-f(r)\right\}\bm{1}_{D_{\partial}}(r) 
				\right].
			\end{align}
			By substituting $f(r)=\delta_{r,r'}$ with the Kronecker delta $\delta_{r,r'}$, 
			the right-hand side can be transformed into 
			\begin{align}
				& \frac{\lambda}{2}\left[
				 P_t(r'-l)\bm{1}_{D_+}(r'-l) - P_t(r')\bm{1}_{D_+}(r')
				+P_t(r'+l)\bm{1}_{D_-}(r'+l) - P_t(r')\bm{1}_{D_-}(r')
				-P_t(r')\bm{1}_{D_{\partial}}(r')
			\right] \notag \\
			&+\frac{\lambda}{2}\delta_{0,r'}\sum_{r\in D_{\partial}}P_t(r).
			\end{align}
			After replacing $r'\to r$, we thus obtain the ML equation
			\begin{align}
				\frac{\partial P_t(r)}{\partial t} 
				=&\frac{\lambda}{2}\left\{
					P_t(r-l)\bm{1}_{D_+}(r-l) - P_t(r)\bm{1}_{D_+}(r)
				+ P_t(r+l)\bm{1}_{D_-}(r+l) - P_t(r)\bm{1}_{D_-}(r)
				-P_t(r)\bm{1}_{D_{\partial}}(r)
				\right\}\notag \\
				&+\frac{\lambda}{2}\delta_{0,r}\left\{P_t(-l\nbar+1)+P_t(l\nbar-l)\right\}.
			\end{align}
			This ML equation can be rewritten as 
			\begin{align}
				\frac{\partial P_t(r)}{\partial t} =& 
					\frac{\lambda}{2}\Delta^2 P_t(r) + 
				\begin{cases} 
					0, & \mbox{if }l \leq |r| \leq l(\nbar-1) \\
					\frac{\lambda}{2}\left\{P_t(-l\nbar+l)+P_t(l\nbar-l)\right\},&  \mbox{if }r = 0 \\
					-\frac{\lambda}{2}P_t(l\nbar-l), &  \mbox{if }r = +l\nbar \\
					+\frac{\lambda}{2}P_t(-l\nbar+l), &  \mbox{if }r = -l\nbar \\
				\end{cases} \notag \\
				=& \frac{\lambda}{2}\Delta^2 P_t(r) + \delta_{r,0}\frac{\lambda}{2}\left\{P_t(-l\nbar+l)+P_t(l\nbar-l)\right\} \notag \\
				&- \delta_{r,l\nbar}\frac{\lambda}{2}P_t(l\nbar-l) + \delta_{r,-l\nbar}\frac{\lambda}{2}P_t(-l\nbar+l),
				\label{eq:master_eq_lattice_cases}
			\end{align}
			where we have used the boundary condition\footnote{
				The obvious relation $\partial P_t(\pm l\nbar)/\partial t = 0$
				should hold at the boundary $r=\pm l\nbar$, which is consistent with the ML equation~\eqref{eq:master_eq_lattice_cases} under the boundary condition~\eqref{eq:master_eq_lattice_boundaryCondition}.
			}
			\begin{equation}
				P_t(r) = 0 \>\> \mbox{ for }|r|\geq l\nbar 
				\label{eq:master_eq_lattice_boundaryCondition}
			\end{equation}
			and the second-order difference operator $\Delta^2$ defined by 
			\begin{equation}
				\Delta^2 P_t(r) := P_t(r+l)-2P_t(r)+P_t(r-l).
			\end{equation}

		\subsubsection{Continuous limit}
			Here we take a continuous limit of the lattice model~\eqref{eq:SDE_discrete} consistently with the original continuous version~\eqref{eq:r_varTrans} according to the system size expansion~\cite{vanKampenB,GardinerB}. In other words, we consider the diffusive limit: 
			\begin{equation}
				l\to 0, \>\>\> \nbar\to \infty, \>\>\> \lambda \to \infty,\>\>\> \lambda l^2 = \sigmaCM^2 = \mbox{const.}, \>\>\> l\nbar = L/2 = \mbox{const.}
				\label{eq:diffusiveLimit}
			\end{equation} 
			The discrete ML equation~\eqref{eq:master_eq_lattice_cases} reduces the continuous version~\eqref{eq:masterEq1_confined} for the diffusive limit~\eqref{eq:diffusiveLimit}. Indeed, by using the Kramers-Moyal expansion~\cite{vanKampenB,GardinerB}
			\begin{align}
				\Delta^2 P_t(r) &\simeq l^2\frac{\partial^2P_t(r)}{\partial r^2} + O(l^3), \\
				P_t(+l\nbar-l) &= P_t(+L/2-l) \simeq P_t(+L/2) - l\partial_{-1}P_t(+L/2) + O(l^2) 
				= - l\partial_{-1}P_t(+L/2) + O(l^2), \\
				P_t(-l\nbar+l) &= P_t(-L/2+l) \simeq P_t(-L/2) + l\partial_{+1}P_t(-L/2) + O(l^2) 
				= + l\partial_{+1}P_t(-L/2) + O(l^2),
			\end{align}
			and replacing the Kronecker $\delta$ with the Dirac $\delta$ function 
			\begin{equation}
				\lim_{l\downarrow 0}\frac{1}{l}\delta_{r,x} = \delta(r-x),
			\end{equation}
			we obtain Eq.~\eqref{eq:masterEq1_confined} from Eq.~\eqref{eq:master_eq_lattice_cases}. This derivation is rather elementary without requiring neither Novikov's theorem nor the functional calculus. We believe that this derivation will develop the better intuition of the general audience. 

\section{Result 2: exact solution to the reduced master-Liouville equation~\eqref{eq:masterEq1_confined}}
	\label{sec:exact_sol_reducedML}
	The ML equation~\eqref{eq:masterEq1_confined} can be solved exactly in the steady state. Let us assume the symmetry of the PDF in the steady state:
	\begin{equation}
		\phi(r):= \lim_{t\to \infty}P_{t}(r), \>\>\> 
		\phi(r) = \phi(-r), \>\>\> 
		\frac{\partial}{\partial r}\phi(r) = -\frac{\partial}{\partial r}\phi(-r).
	\end{equation}
	This symmetry implies that it is sufficient to investigate the steady solution $\phi(r)$ only for $r\in [0,\infty)$. On the basis of the ML equation~\eqref{eq:masterEq1_confined}, $\phi(r)$ satisfies 
	\begin{equation}
		\frac{\sigmaCM^2}{2}\frac{\partial^2}{\partial r^2}\phi(r) = 0, \>\>\>\>\> r\in (0,L/2), 
	\end{equation}
	which implies the piecewise-linear steady solution
	\begin{equation}
		\phi(r) = C_1 + C_2 r, \>\>\>\>\> r\in (0,L/2)
		\label{eq:piecewise-linear_steadySolution}
	\end{equation}
	with coefficients $C_1$ and $C_2$. We will then consider the boundary condition at $r=0, L/2$.

	\subsection{Boundary condition at $r=0$}
		The boundary condition at $r=0$ is given by integrating Eq.~\eqref{eq:masterEq1_confined} over $(-h,+h)$ with infinitesimal positive $h>0$ as 
		\begin{align}
			0 &= \int_{-h}^{+h} dr \left[\frac{\sigmaCM^2}{2}\frac{\partial^2}{\partial r^2}\phi(r) - \delta(r)\frac{\sigmaCM^2}{2}\partial_{-1}\phi(L/2)+ \delta(r)\frac{\sigmaCM^2}{2}\partial_{+1}\phi(-L/2)\right] \notag \\ 
			&= \frac{\sigmaCM^2}{2}\left(\frac{\partial}{\partial r}\phi(h)-\frac{\partial}{\partial r}\phi(-h)\right) - \frac{\sigmaCM^2}{2}\partial_{-1}\phi(L/2) + \frac{\sigmaCM^2}{2}\partial_{+1}\phi(-L/2) \notag \\
			&= \sigmaCM^2\frac{\partial}{\partial r}\phi(h) - \sigmaCM^2\partial_{-1}\phi(L/2),
		\end{align}
		where we have used $|\partial_{-s}\phi(sL/2)|=-s\partial_{-s}\phi(sL/2)$. For $h\downarrow 0$, we obtain the boundary condition 
		\begin{equation}
			\partial_{+1} \phi(0) = \partial_{-1}\phi(L/2). 
			\label{eq:boundary_condition_r=0}
		\end{equation}

	\subsection{Boundary condition at $r=L/2$}
		Here we examine the boundary condition at $r=L/2$. Let us integrate the ML equation~\eqref{eq:masterEq1_confined} over $(L/2-h,L/2+h)$ with infinitesimal positive $h>0$ as 
		\begin{align}
			0	= &\int_{L/2-h}^{L/2+h} dr \left[\frac{\sigmaCM^2}{2}\frac{\partial^2}{\partial r^2}\phi(r) + \delta(r-L/2)\frac{\sigmaCM^2}{2}\left|\partial_{-1}\phi(L/2)\right|\right] \notag \\
			= &\frac{\sigmaCM^2}{2}\frac{\partial}{\partial r}\phi(L/2-h) - \frac{\sigmaCM^2}{2}\frac{\partial}{\partial r}\phi(L/2+h) + \frac{\sigmaCM^2}{2}\left|\partial_{-1}\phi(L/2)\right|.
		\end{align}
		Considering the fact $\partial_{-1}\phi(L/2) < 0$, we obtain 
		\begin{equation}
				\frac{\partial}{\partial r}\phi(L/2+h) = \frac{\partial}{\partial r}\phi(L/2-h) - \partial_{-1}\phi(L/2).
		\end{equation}
		By taking the limit $h\to 0$, we obtain the boundary condition 
		\begin{equation}
			\lim_{h\downarrow 0} \frac{\partial}{\partial r}\phi(L/2+h) = \partial_{+1}\phi(L/2) = 0.
			\label{eq:boundary_condition_anytime_r=L}
		\end{equation}

	\subsection{Normalisation condition and the explicit steady solution}
		Considering the boundary conditions~\eqref{eq:boundary_condition_r=0} and \eqref{eq:boundary_condition_anytime_r=L} together with the local solution~\eqref{eq:piecewise-linear_steadySolution} for $r\in(0,L/2)$, the global solution is given by 
		\begin{equation}
			\phi(r) = 
			\begin{cases}
				C_1 + C_2 r & (r\in (0,L/2)) \\
				C_1 + C_2 L/2 = {\rm const.} & (r\in [L/2,\infty)) 
			\end{cases}.
		\end{equation}
		The coefficients $C_1$ and $C_2$ for the steady solution~\eqref{eq:piecewise-linear_steadySolution} is determined by the normalisation condition: 
		\begin{equation}
			\int_{-\infty}^{\infty} dr \phi(r) = 2\int_{0}^{\infty} dr \phi(r) = 1,
		\end{equation}
		which deduces the tent function (see Fig.~\ref{fig:tent-functions}a)
		\begin{equation}
			\phi(r) = \max\left\{0, \frac{L/2-|r|}{L^2/4}\right\}
			\label{eq:tent-function_SS}
		\end{equation}
		with $C_1 = 2/L$ and $C_2 = -4/L^2$. 
	
	\subsection{Average order-book profile}
		\begin{figure}
			\centering 
			\includegraphics[width=150mm]{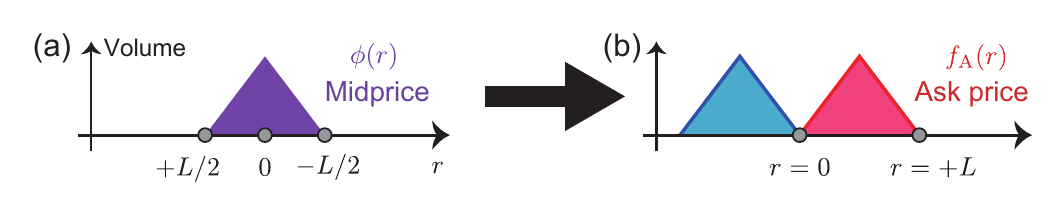}
			\caption{
				(a)~The average PDF $\phi(r):=\lim_{t\to \infty}P_t(r)$ in the steady state is given by the tent function~\eqref{eq:tent-function_SS}. (b)~$\phi(r)$ is directly related to the average order-book profile~\eqref{eq:tent-function_SS_OB}, where the depth $r$ is measured from the centre of mass $\hz_{\CM}$. 
			}
			\label{fig:tent-functions}
		\end{figure}
		Equation~\eqref{eq:tent-function_SS} is directly related to the normalised average order-book profile $f_{A}(r)$ (see Fig.~\ref{fig:tent-functions}b) as 
		\begin{equation}
			f_{\mrA}(r) := \left< \frac{1}{2}\sum_{i=1,2}\delta\left(\ha_i - \hz_{\CM} - r\right)\right> = \phi(r-L/2) = \max\left\{0, \frac{L/2-|r-L/2|}{L^2/4}\right\},
			\label{eq:tent-function_SS_OB}
		\end{equation}
		where the depth $r$ is measured from the centre of mass $\hz_{\CM}$, or equivalently from the market midprice for the special case of $N=2$.

	\subsection{Average transaction interval}
		Here we discuss the average transaction interval. The average transaction interval can be evaluated by considering the physical meaning of the steady probability current $j_{\mrss}(r):= \lim_{t\to \infty}j_t(r)$. Indeed, since the absolute value of the steady probability current $|j_{\mrss}(r)|$ at the walls $r=\pm L/2$ represents the transaction probability per unit time, the average transaction interval $\la\tau \ra$ is given by
		\begin{equation}
			\la \tau \ra = \lim_{h\downarrow 0}\frac{1}{|j_{\mrss}(-L/2+h)| + |j_{\mrss}(L/2-h)|} = \frac{L^2}{4\sigmaCM^2}= \frac{L^2}{2\sigma^2}.
		\end{equation}
		This is exactly equal to the formula~\eqref{eq:review_YamadaPRE2009_<tau>} derived in Ref.~\cite{YamadaPRE2009}. We thus rederive the statistics of the transaction interval via the kinetic theory. 

\section{Result 3: the full master-Liouville equation}
	\label{sec:fullML_der}
	We have derived the reduced ML equation to exactly obtain the average order-book profile and the average transaction interval. In this section, we derive the full ML equation for the two-body dealer model~\eqref{eq:dealermodel} via Novikov's theorem	in the parallel calculation to Sec.~\ref{sec:derMaster-reduced1_Novikov}:
	\begin{screen}
		\begin{subequations}
			\label{eq:fullMasterEq}
			The full PDF $P_t(z_1,z_2):=\la \delta(z_1-\hz_1)\delta(z_2-\hz_2)\ra$ obeys the {\it full ML equation} defined by
			\begin{equation}
				\frac{\partial P_t(z_1,z_2)}{\partial t} = \sum_{i=1,2}\frac{\sigma^2}{2} \frac{\partial^2 P_t(z_1,z_2)}{\partial z_i^2} + \sum_{s=\pm 1}\left[J_{t;s}(z_1+sL/2,z_2-sL/2)-J_{t;s}(z_1,z_2)\right]
			\end{equation}
			with the {\it probability current} $J_{t;s}(z_1,z_2)$ defined by
			\begin{equation}
				J_{t;s}(z_1,z_2):=\frac{-s\sigma^2}{2}\delta(z_1-z_2-sL)\tl{\partial}_{12;s}P_t\left(z_1,z_2\right) = \frac{\sigma^2}{2}\delta(z_1-z_2-sL)\left|\tl{\partial}_{12;s}\right|P_t\left(z_1,z_2\right) \geq 0.
			\end{equation}
		\end{subequations}
	\end{screen}
	Here we have introduced the left ($s=-1$) and right ($s=+1$) derivatives defined by
	\begin{equation}
			\partial_{1;s}f(z_1,z_2):= \lim_{h\downarrow 0}\frac{f(z_1+sh,z_2)-f(z_1,z_2)}{sh}, \>\>\> 
			\partial_{2;s}f(z_1,z_2):= \lim_{h\downarrow 0}\frac{f(z_1,z_2+sh)-f(z_1,z_2)}{sh},
	\end{equation} 
	and 
	\begin{align}
		\tl{\partial}_{12;s}f(z_1,z_2) &:= \partial_{1;-s}f(z_1,z_2) - \partial_{2;s}f(z_1,z_2), \\
		\left|\tl{\partial}_{12;s}\right|f(z_1,z_2) &:= \left|\partial_{1;-s}f(z_1,z_2)\right| + \left|\partial_{2;s}f(z_1,z_2)\right| \geq 0. 
	\end{align}
	Finally, we confirm that the full ML equation~\eqref{eq:fullMasterEq} is consistent with the reduced ML equation~\eqref{eq:masterEq1_confined}. 
	
	\subsection{Reformulation based on the OU coloured noise}
		First, we reformulate the SDE~\eqref{eq:dealermodel} using the OU coloured noise: 
		\begin{subequations}
			\label{eq:SDE_full_trans}
			\begin{align}
				\hz_1(t+dt) &= 	\hz_1(t) +
				\begin{cases}
					\sigma \heta_{1;\eps}(t)dt & \mbox{if $|\hz_1(t)-\hz_2(t)|<L$} \\
					-L/2 & \mbox{if $\hz_1(t)-\hz_2(t)=+L$} \\
					+L/2 & \mbox{if $\hz_1(t)-\hz_2(t)=-L$}
				\end{cases} \\
				\hz_2(t+dt) &= \hz_2(t) + 
				\begin{cases}
					\sigma \heta_{2;\eps}(t)dt & \mbox{if $|\hz_1(t)-\hz_2(t)|<L$} \\
					+L/2 & \mbox{if $\hz_1(t)-\hz_2(t)=+L$} \\
					-L/2 & \mbox{if $\hz_1(t)-\hz_2(t)=-L$}
				\end{cases}
			\end{align}
			where the OU coloured noises $\heta_{1;\eps}$ and $\heta_{2;\eps}$ are defined by
			\begin{equation}
				\frac{d\heta_{1;\eps}}{dt} = -\frac{1}{\eps}\heta_{1;\eps} + \hxi^{\mrG}_1, \>\>\>
				\frac{d\heta_{2;\eps}}{dt} = -\frac{1}{\eps}\heta_{2;\eps} + \hxi^{\mrG}_2,
			\end{equation}
			where $\hxi^{\mrG}_1$ and $\hxi^{\mrG}_2$ are the standard independent white Gaussian noises (i.e., $\la \hxi^{\mrG}_i\ra=0$ and $\la \hxi^{\mrG}_i(t_k)\hxi^{\mrG}_j(t_l) \ra = \delta_{i,j}\delta(t_k-t_l)$). We finally take the white-noise limit $\eps\downarrow 0$ to keep the consistency with the original model~\eqref{eq:dealermodel}. 
		\end{subequations}

		Equations~\eqref{eq:SDE_full_trans} can be rewritten by the $\delta$ functions as follows: let us introduce the transaction time $\htau_{s;i}$ as the $i$th transaction time with sign $s=\pm 1$ satisfying
		\begin{subequations}
			\label{eq:SDE_full_delta}
			\begin{equation}
				\hz_1(\htau_{s;i})-\hz_2(\htau_{s;i})=sL, \>\>\>\htau_{s;i}<\htau_{s;i+1}, \>\>\> s\in \{-1,+1\}.
			\end{equation}
			Using the $\delta$ functions, we can rewrite Eq.~\eqref{eq:SDE_full_trans} as 
			\begin{align}
				\frac{d\hz_1}{dt} &= \sigma \heta_{1;\eps}(t)dt - \sum_{s=\pm 1}\sum_{i} \frac{sL}{2}\delta(t-\htau_{s;i}), \\
				\frac{d\hz_2}{dt} &= \sigma \heta_{2;\eps}(t)dt + \sum_{s=\pm 1}\sum_{i} \frac{sL}{2}\delta(t-\htau_{s;i}).
			\end{align}
		\end{subequations}

		We note that, for the collision $\hz_1(\htau_{s;i})-\hz_2(\htau_{s;i})=sL$, the relative velocity $d(\hz_1-\hz_2)/dt$ must be positive for $s=1$ and negative for $s=-1$, respectively. In other words, the following relation holds, 
		\begin{equation}
			s\lim_{h\downarrow 0}\frac{d}{dt}\left\{\hz_1(\htau_{s;i}-h)-\hz_2(\htau_{s;i}-h)\right\}
			= s\sigma\lim_{h\downarrow 0}\left\{\heta_{1;\eps}(\htau_{s;i}-h)-\heta_{2;\eps}(\htau_{s;i}-h)\right\} > 0.
			\label{eq:sign_relative_velocity_fullMaster}
		\end{equation}

	\subsection{Dynamics of an arbitrary function $f(\hz_1,\hz_2)$}
		We next consider the dynamics of an arbitrary function $f(\hz_1,\hz_2)$: 
		\begin{align}
			\frac{df(\hz_1,\hz_2)}{dt} =& 
				\sigma\heta_{1;\eps}\frac{\partial f(\hz_1,\hz_2)}{\partial \hz_1} + 
				\sigma\heta_{2;\eps}\frac{\partial f(\hz_1,\hz_2)}{\partial \hz_2} \notag \\
				&+\sum_{s=\pm 1}\sum_i \left[f(\hz_1-sL/2,\hz_2+sL/2)-f(\hz_1,\hz_2)\right]\delta(t-\htau_{s;i}). 
				\label{eq:fullmasterEq_trans_2}
		\end{align}
		By the way, since $\htau_{s;i}$ is the solution of $\hz_1(\htau_{s;i})-\hz_2(\htau_{s;i})=sL$, the $\delta$ function $\delta(\hz_1-\hz_2-sL)$ can be decomposed as follows: 
		\begin{equation}
			\delta(\hz_1-\hz_2-sL) = \sum_{i} \left|\frac{d}{dt}\left(\hz_1-\hz_2-sL\right)\right|^{-1}\delta(t-\htau_{s;i})
			= \sum_{i} \sigma^{-1}\left|\heta_{1;\eps}-\heta_{2;\eps}\right|^{-1}\delta(t-\htau_{s;i})
		\end{equation}
		By considering the sign of the relative velocity (equivalently, the direction of collisions) given by Eq.~\eqref{eq:sign_relative_velocity_fullMaster}, we obtain 
		\begin{equation}
			s\sigma g(\hz_1,\hz_2)(\heta_{1;\eps}-\heta_{2;\eps})\delta(\hz_1-\hz_2-sL) = 
			\sum_{i}sg(\hz_1,\hz_2)\delta(t-\htau_{s;i})
		\end{equation} 
		for an arbitrary function $g(\hz_1,\hz_2)$. Equation~\eqref{eq:fullmasterEq_trans_2} then can be rewritten as 
		\begin{align}
			\frac{df(\hz_1,\hz_2)}{dt} =& 
				\sigma\heta_{1;\eps}\frac{\partial f(\hz_1,\hz_2)}{\partial \hz_1} + 
				\sigma\heta_{2;\eps}\frac{\partial f(\hz_1,\hz_2)}{\partial \hz_2} \notag \\
				&+\sum_{s=\pm 1}s\sigma (\heta_{1;\eps}-\heta_{2;\eps})\left[f(\hz_1-sL/2,\hz_2+sL/2)-f(\hz_1,\hz_2)\right]\delta(\hz_1-\hz_2-sL). 
		\end{align}
		We then take the ensemble average of both sides as 
		\begin{align}
			\left<\frac{df(\hz_1,\hz_2)}{dt}\right> =& 
			\left<\sigma\heta_{1;\eps}\frac{\partial f(\hz_1,\hz_2)}{\partial \hz_1} + 
				\sigma\heta_{2;\eps}\frac{\partial f(\hz_1,\hz_2)}{\partial \hz_2}\right> \notag \\
				&+\sum_{s=\pm 1}s\sigma\left< (\heta_{1;\eps}-\heta_{2;\eps})\left[f(\hz_1-sL/2,\hz_2+sL/2)-f(\hz_1,\hz_2)\right]\delta(\hz_1-\hz_2-sL)\right>. 
				\label{eq:fullmasterEq_trans_3}
		\end{align}

	\subsection{Application of Novikov's theorem}
		To proceed the calculation further, we have to calculate the correlation terms of the form $\la \heta_{i;\eps}(t)g(\hz_1,\hz_2)\ra$. Such a correlation term can be calculated via Novikov's theorem, 
		\begin{align}
			\la \heta_{i;\eps}(t)g(\hz_1,\hz_2)\ra 
			&= \int_{0}^t dt'\la\heta_{i;\eps}(t)\heta_{i;\eps}(t')\ra \left<\frac{\delta g(\hz_1,\hz_2)}{\delta \heta_{i;\eps}}\right> \notag \\
			&= \int_{0}^t dt'\frac{e^{-(t-t')/\eps}}{2\eps} \left<\left\{
				\frac{\delta \hz_1(t)}{\delta \heta_{i;\eps}(t')}\frac{\partial}{\partial \hz_1} +
				\frac{\delta \hz_2(t)}{\delta \heta_{i;\eps}(t')}\frac{\partial}{\partial \hz_2} 
					\right\}g(\hz_1,\hz_2)\right>,
		\end{align}
		which implies 
		\begin{equation}
			\lim_{\eps\downarrow 0} \la \heta_{i;\eps}(t)g(\hz_1,\hz_2)\ra 
			= \lim_{t'\uparrow t}	\frac{1}{2}\left<\left\{
				\frac{\delta \hz_1(t)}{\delta \heta_{i;\eps}(t')}\frac{\partial}{\partial \hz_1} +
				\frac{\delta \hz_2(t)}{\delta \heta_{i;\eps}(t')}\frac{\partial}{\partial \hz_2} 
					\right\}g(\hz_1,\hz_2)\right>
		\end{equation}
		for the white-noise limit. Since the formal solution of the SDE~\eqref{eq:SDE_full_delta} is given by 
		\begin{align}
			\hz_1(t) &= \hz_1(t_{\rm ini}) + \sigma \int_{t_{\rm ini}}^t dt' \heta_{1;\eps}(t'), \\
			\hz_2(t) &= \hz_2(t_{\rm ini}) + \sigma \int_{t_{\rm ini}}^t dt' \heta_{2;\eps}(t'),
		\end{align}
		assuming the absense of transactions during $[t_{\rm ini},t)$, 
		we obtain 
		\begin{equation}
			\lim_{t'\uparrow t}\frac{\delta \hz_i(t)}{\delta \heta_{j}(t')} = \sigma.
		\end{equation}
		We thus obtain 
		\begin{equation}
			\lim_{\eps\downarrow 0} \la \heta_{i;\eps}(t)g(\hz_1,\hz_2)\ra 
			= \frac{\sigma}{2}\left<\frac{\partial}{\partial \hz_i}g(\hz_1,\hz_2)\right>. 
			\label{eq:Novikov_full_2body_fin}
		\end{equation}

		The formula~\eqref{eq:Novikov_full_2body_fin} is useful in deriving the full ML equation. Indeed, for the white-noise limit $\eps\downarrow 0$, we obtain
		\begin{equation}
			\left<\frac{df(\hz_1,\hz_2)}{dt}\right> = \int_{-\infty}^\infty dz_1dz_2 \frac{\partial P_t(z_1,z_2)}{\partial t}f(z_1,z_2),
		\end{equation}
		\begin{equation}
			\left<\sigma\heta_{i;\eps}\frac{\partial f(\hz_1,\hz_2)}{\partial \hz_i}\right>
			= \int_{-\infty}^\infty dz_1dz_2 P_t(z_1,z_2)\frac{\sigma^2}{2} 
			\frac{\partial^2 f(\hz_1,\hz_2)}{\partial z_i^2}
			= \int_{-\infty}^\infty dz_1dz_2 f(\hz_1,\hz_2)\frac{\sigma^2}{2} 
			\frac{\partial^2 P_t(z_1,z_2)}{\partial z_i^2},
		\end{equation}
		\begin{align}
			&\left< (\heta_{1;\eps}-\heta_{2;\eps})f(\hz_1-sL/2,\hz_2+sL/2)\delta(\hz_1-\hz_2-sL)\right> \notag \\ 
			=& \frac{\sigma}{2}\int_{-\infty}^\infty dz_1dz_2 P_t(z_1,z_2) \left(\frac{\partial}{\partial z_1}-\frac{\partial}{\partial z_2}\right)f(z_1-sL/2,z_2+sL/2)\delta(z_1-z_2-sL) \notag \\
			=& -\frac{\sigma}{2} \int_{-\infty}^\infty dz_1dz_2 f(z_1-sL/2,z_2+sL/2)\delta(z_1-z_2-sL)\left(\frac{\partial}{\partial z_1}-\frac{\partial}{\partial z_2}\right)P_t(z_1,z_2),
		\end{align}
		and
		\begin{align}
			&\left< (\heta_{1;\eps}-\heta_{2;\eps})f(\hz_1,\hz_2)\delta(\hz_1-\hz_2-sL)\right> \notag \\
			=& \frac{\sigma}{2}\int_{-\infty}^\infty dz_1dz_2 P_t(z_1,z_2) \left(\frac{\partial}{\partial z_1}-\frac{\partial}{\partial z_2}\right)f(z_1,z_2)\delta(z_1-z_2-sL) \notag \\
			=& -\frac{\sigma}{2} \int_{-\infty}^\infty dz_1dz_2 f(z_1,z_2)\delta(z_1-z_2-sL)\left(\frac{\partial}{\partial z_1}-\frac{\partial}{\partial z_2}\right)P_t(z_1,z_2),
		\end{align}
		where we have performed the partial integration. By introducing a symbol
		\begin{equation}
			\tl{\partial}_{12} := \frac{\partial}{\partial z_1}-\frac{\partial}{\partial z_2}
		\end{equation}
		and by substituting $f(\hz_1,\hz_2)=\delta(\hz_1-z_1)\delta (\hz_2-z_2)$, we obtain the full ML equation
		\begin{align}
			\frac{\partial P_t(z_1,z_2)}{\partial t} = \sum_{i=1,2}\frac{\sigma^2}{2} \frac{\partial^2 P_t(z_1,z_2)}{\partial z_i^2}
			+&\sum_{s=\pm 1}\frac{-s\sigma^2}{2}\delta(z_1-z_2)\tl{\partial}_{12}P_t\left(z_1+\frac{sL}{2},z_2-\frac{sL}{2}\right)\notag \\
			-&\sum_{s=\pm 1}\frac{-s\sigma^2}{2}\delta(z_1-z_2-sL)\tl{\partial}_{12}P_t(z_1,z_2).
			\label{eq:full_master_trans6}
		\end{align}
		We will further rewrite this ML equation by considering several technical issues. 

	\subsection{Technical issues on the left and right derivatives}
		Here we consider the technical issues on the left and right derivatives in Eq.~\eqref{eq:full_master_trans6}. Let us first consider the meaning of the derivatives in $\tl{\partial}_{12}$ for the contribution of $s=1$. For the nonnegativity of the probability, 
		\begin{subequations}
			\label{ineq:prob_zeroOrPositive_1_fullMaster}
			\begin{equation}
				P_t(z_1,z_2) \geq 0 \>\>\> \mbox{ for all $z_1,z_2$}.  
			\end{equation}
			At the same time, the transaction rule imposes an obvious restriction, 
			\begin{equation}
				P_t(z_1,z_2) = 0 \>\>\> \mbox{ for $z_1-z_2 \geq L$}.  
			\end{equation}
		\end{subequations}
		\begin{figure}
			\centering
			\includegraphics[width=130mm]{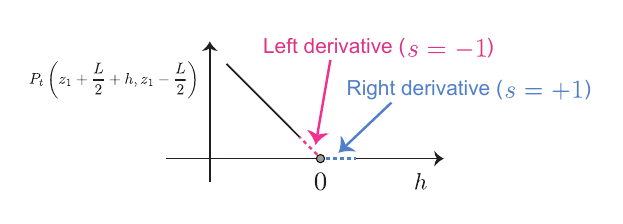}
			\caption{
				Schematic of the left ($\partial_{1;-1}$) and right ($\partial_{1;+1}$) derivatives of $P_t(z_1+L/2,z_2-L/2)$ on the condition $z_2=z_1$. The right derivative is zero, while the left is nonpositive.
			}
			\label{fig:SignOfDerivative_full}
		\end{figure}
		As illustrated in Fig.~\ref{fig:SignOfDerivative_full}, this implies that 
		\begin{subequations}
			\label{eq_zero_trans10}
			\begin{align}
				\partial_{1;+1}P_t\left(z_1+\frac{L}{2},z_1-\frac{L}{2}\right):=& \lim_{h\downarrow 0}\frac{P_t\left(z_1+\frac{L}{2}+h,z_1-\frac{L}{2}\right)-P_t\left(z_1+\frac{L}{2},z_1-\frac{L}{2}\right)}{h} = \lim_{h\downarrow 0}\frac{0-0}{h} = 0, \\
				\partial_{2;-1}P_t\left(z_1+\frac{L}{2},z_1-\frac{L}{2}\right):=& \lim_{h\downarrow 0}\frac{P_t\left(z_1+\frac{L}{2},z_1-\frac{L}{2}-h\right)-P_t\left(z_1+\frac{L}{2},z_1-\frac{L}{2}\right)}{-h} = \lim_{h\downarrow 0}\frac{0-0}{-h} = 0,
			\end{align}
		\end{subequations}
		where we have introduced the left ($s=-1$) and right ($s=+1$) derivatives as
		\begin{equation}
			\partial_{1;s}f(z_1,z_2):= \lim_{h\downarrow 0}\frac{f(z_1+sh,z_2)-f(z_1,z_2)}{sh}, \>\>\> 
			\partial_{2;s}f(z_1,z_2):= \lim_{h\downarrow 0}\frac{f(z_1,z_2+sh)-f(z_1,z_2)}{sh}.
		\end{equation}
		We can also show 
		\begin{equation}
			\label{eq_zero_trans11}
			\partial_{1;+1}P_t\left(z_1,z_1-L\right)=\partial_{2;-1}P_t\left(z_1,z_1-L\right)=0.
		\end{equation}
		Considering the relations~\eqref{eq_zero_trans10} and~\eqref{eq_zero_trans11}, the derivatives $\tl{\partial}_{12}$ for $s=1$ should be understood as 
		\begin{equation}
			\tl{\partial}_{12} \to \tl{\partial}_{12;+1} := \partial_{1;-1} - \partial_{2;+1}.
		\end{equation}	
		Similarly, by considering the apparent restriction imposed by the transaction rule 
		\begin{equation}
			P_t(z_1,z_2) = 0 \>\>\> \mbox{ for $z_1-z_2 \leq -L$}, 
		\end{equation}
		the derivative $\tl{\partial}_{12}$ for $s=-1$ should be understood as 
		\begin{equation}
			\tl{\partial}_{12} \to \tl{\partial}_{12;-1} := \partial_{1;+1} - \partial_{2;-1}.
		\end{equation}
		In summary, the ML equation~\eqref{eq:full_master_trans6} should be technically interpreted as 
		\begin{align}
			\frac{\partial P_t(z_1,z_2)}{\partial t} = \sum_{i=1,2}\frac{\sigma^2}{2} \frac{\partial^2 P_t(z_1,z_2)}{\partial z_i^2}
			+&\sum_{s=\pm 1}\frac{-s\sigma^2}{2}\delta(z_1-z_2)\tl{\partial}_{12;s}P_t\left(z_1+\frac{sL}{2},z_2-\frac{sL}{2}\right)\notag \\
			-&\sum_{s=\pm 1}\frac{-s\sigma^2}{2}\delta(z_1-z_2-sL)\tl{\partial}_{12;s}P_t(z_1,z_2)
			\label{eq:full_master_trans7}
		\end{align}
		in terms of the left and right derivatives. 

	\subsection{Sign of derivatives}
		Next, we consider the sign of the derivatives. As illustrated in Fig.~\ref{fig:SignOfDerivative_full}, the relation~\eqref{ineq:prob_zeroOrPositive_1_fullMaster} implies 
		\begin{subequations}
			\begin{align}
				\partial_{1;-1}P_t\left(z_1+\frac{L}{2},z_1-\frac{L}{2}\right):=& \lim_{h\downarrow 0}\frac{P_t\left(z_1+\frac{L}{2}-h,z_1-\frac{L}{2}\right)-P_t\left(z_1+\frac{L}{2},z_1-\frac{L}{2}\right)}{-h} \notag \\
				=& \lim_{h\downarrow 0}\frac{P_t\left(z_1+\frac{L}{2}-h,z_1-\frac{L}{2}\right)-0}{-h} \leq 0, \\
				\partial_{2;+1}P_t\left(z_1+\frac{L}{2},z_1-\frac{L}{2}\right):=& \lim_{h\downarrow 0}\frac{P_t\left(z_1+\frac{L}{2},z_1-\frac{L}{2}+h\right)-P_t\left(z_1+\frac{L}{2},z_1-\frac{L}{2}\right)}{h} \notag \\
				=& \lim_{h\downarrow 0}\frac{P_t\left(z_1+\frac{L}{2},z_1-\frac{L}{2}+h\right)-0}{h} \geq 0. 
			\end{align}
		\end{subequations}
		We can also show 
		\begin{equation}
			\partial_{1;-1}P_t\left(z_1,z_1-L\right)\leq 0, \>\>\> 
			\partial_{2;+1}P_t\left(z_1,z_1-L\right)\geq 0. 
		\end{equation}
		This implies that 
		\begin{subequations}
			\label{ineq:prob_flux_21}
			\begin{align}
				\frac{-s\sigma^2}{2}\delta(z_1-z_2)\tl{\partial}_{12;s}P_t\left(z_1+\frac{sL}{2},z_2-\frac{sL}{2}\right) &\geq 0, \\
				\frac{-s\sigma^2}{2}\delta(z_1-z_2-sL)\tl{\partial}_{12;s}P_t\left(z_1,z_2\right) &\geq 0
			\end{align}
		\end{subequations}
		for $s=+1$. Similarly, we can show the inequality~\eqref{ineq:prob_flux_21} even for $s=-1$. By considering the inequality~\eqref{ineq:prob_flux_21}, it is useful to introduce the nonnegative probability current: 
		\begin{equation}
			J_{t;s}(z_1,z_2):=\frac{-s\sigma^2}{2}\delta(z_1-z_2-sL)\tl{\partial}_{12;s}P_t\left(z_1,z_2\right) \geq 0. 
		\end{equation}
		Using the nonnegative probability current $J_{t;s}(z_1,z_2)$, we can rewrite the ML equation~\eqref{eq:full_master_trans7} as 
		\begin{equation}
			\frac{\partial P_t(z_1,z_2)}{\partial t} = \sum_{i=1,2}\frac{\sigma^2}{2} \frac{\partial^2 P_t(z_1,z_2)}{\partial z_i^2} + \sum_{s=\pm 1}\left[J_{t;s}(z_1+sL/2,z_2-sL/2)-J_{t;s}(z_1,z_2)\right].
		\end{equation}
		
		Considering the nonnegativity of the probability current $J_{t;s}(z_1,z_2)\geq 0$, it would be useful to introduce a notation where the nonnegativity is apparent. We thus introduce the symbol $|\tl{\partial}_{12;s}|$ defined by
		\begin{equation}
			\left|\tl{\partial}_{12;s}\right|f(z_1,z_2) := \left|\partial_{1;-s}f(z_1,z_2)\right| + \left|\partial_{2;s}f(z_1,z_2)\right| \geq 0,
		\end{equation}
		whose nonnegativity is apparent by its definition. We can thus rewrite the probability current $J_{t;s}(z_1,z_2)$ as 
		\begin{equation}
			J_{t;s}(z_1,z_2):=\frac{\sigma^2}{2}\delta(z_1-z_2-sL)\left|\tl{\partial}_{12;s}\right|P_t\left(z_1,z_2\right) \geq 0.
		\end{equation}
		We thus obtain the full ML equation~\eqref{eq:fullMasterEq}. 
		
		Notably, the selection of the left and right derivatives in the ML equations were not apparent in our previous publications~\cite{KanazawaPRL2018,KanazawaPRE2018}. In this sense, the full ML equation~\eqref{eq:fullMasterEq} that we have derived in this report is the complete form in terms of the mathematical interpretation. 
	
	\subsection{Intuitive interpretation of the full master-Liouville equation~\eqref{eq:fullMasterEq}}
		\begin{figure}
			\centering
			\includegraphics[width=155mm]{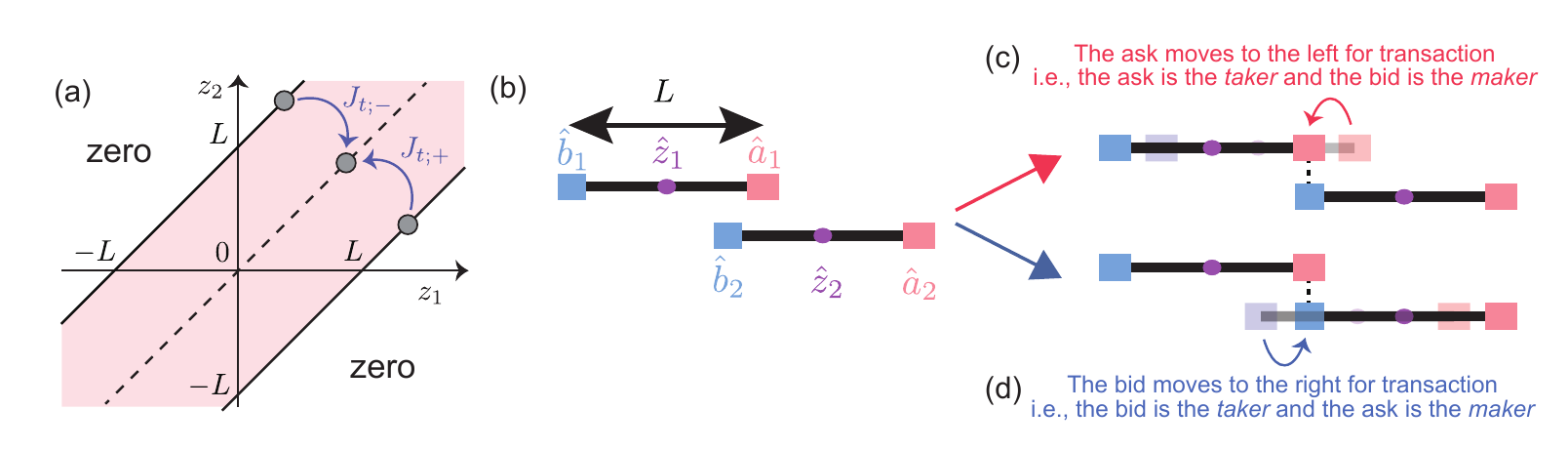}
			\caption{
				(a)~The transaction rule imposes the condition $P_t(z_1,z_2)=0$ for $|z_1-z_2|\geq L$ and the transaction occurs on the line $|z_1-z_2|=L$. In the light-red regime $|z_1-z_2|<L$, the PDF can be non-zero such that $P_t(z_1,z_2)>0$. The full ML equation~\eqref{eq:fullMasterEq} implies that the probability current $J_{t;s}$ on the line $z_1-z_2=sL$ is transferred to the line $z_1=z_2$ for both $s=\pm 1$. 
				(b)~Let us consider the time just before a transaction. There are two scenarios: 
        (c)~The ask price $a_1$ moves to the left, leading to a transaction. This type of transactions implies that the ask is the taker, and the bid is the maker. 
        (d)~The bid price $b_2$ moves to the right, leading to a transaction. This type of transactions implies that the bid is the taker and the ask is the maker. 
			}
			\label{fig:ProbabilityCurrent_full}
		\end{figure}
		Here we provide an intuitive interpretation of the full ML equation~\eqref{eq:fullMasterEq} from the viewpoint of the probability current. In this subsection, we abbreviate the technical symbol of the left and right derivatives for simplicity. The transaction rule apparently imposes the rule 
		\begin{equation}
			P_t(z_1,z_2) = 0 \>\>\> \mbox{ for $|z_1-z_2|\geq L$}
		\end{equation}
		and the transaction occurs on the lines $|z_1-z_2|=L$. The full ML equation~\eqref{eq:fullMasterEq} means that the probability current $J_{t;s}$ on the line $z_1-z_2=sL$ is transferred to the line $z_1=z_2$ for $s=\{+1,-1\}$, due to the transaction rule (see Fig.~\ref{fig:ProbabilityCurrent_full}a). 
	
		In addition, we can intuitive decompose the probability current as 
		\begin{screen}
			\begin{equation}
				J_{t;s}(z_1,z_2) = \delta(z_1-z_2-sL)\sum_{i=1,2}\left|j_{t;s}^{(i)}(z_1,z_2)\right|, \>\>\>
				j_{t;s}^{(i)}(z_1,z_2) := -\frac{\sigma^2}{2}\frac{\partial}{\partial z_i}P_t(z_1,z_2). 
			\end{equation}
		\end{screen}
		The decomposed probability current $j_{t;s}^{(i)}$ represents the transaction where the $i$th trader is the {\it taker} and the other is the {\it maker}. In addition, similarly to Eqs.~\eqref{eq:reducedML_probCurrentForm}, we can rewrite the full ML equation as 
		\begin{screen}
			\begin{equation}
				\frac{\partial P_t(z_1,z_2)}{\partial t} = - \sum_{i=1,2}\frac{\partial }{\partial z_i}j_{t;s}^{(i)}(z_1,z_2) + \sum_{s=\pm 1}\left[J_{t;s}(z_1+sL/2,z_2-sL/2)-J_{t;s}(z_1,z_2)\right].
			\end{equation}
		\end{screen}
		Apparently, this formula is a natural extension of the probability-current representation~\eqref{eq:reducedML_probCurrentForm} of the reduced ML equation. 

		Let us explain the necessary background knowledge on takers and makers in financial markets. In the double-auction systems, the trader leading to the decision via the market orders is called a {\it taker}, whereas the trader waiting for transactions is called {\it maker}. This distinction is practically vital because the takers and makers regarded as liquidity consumers and providers, respectively. Many market regulators offer a financial incentive to makers because sufficient liquidity provision will stabilise the market. 

    From the viewpoints of the taker and maker, any transaction can be classified whether the first trader is a taker or maker. For example, let us consider the timing just before a transaction $\ha_1=\hb_2$ (see Fig.~\ref{fig:ProbabilityCurrent_full}b). There are two classifications for this transaction: the first case is that the ask price $\ha_1$ moves to the left and then leads to the transaction (Fig.~\ref{fig:ProbabilityCurrent_full}c). In this case, the first trader is the taker and the second trader is the maker. Another case is that the bid price moves to the right and then leads to the transaction. In this case, the second trader is the taker, and the first trader is the maker (Fig.~\ref{fig:ProbabilityCurrent_full}d). Considering that the probability current $j_{t;s}^{(i)}$ originates from the diffusion of the $i$th trader, $j_{t;s}^{(i)}$ represents the contribution where the $i$th trader leads a transaction as the taker. 

	\subsection{Consistency confirmation with the reduced master-Liouville equation~\eqref{eq:masterEq1_confined}}
		Finally, let us confirm the consistency between the full and reduced ML equations~\eqref{eq:fullMasterEq} and~\eqref{eq:masterEq1_confined}. In this subsection, the technical symbol on the left and right derivatives are abbreviated for simplicity, such as $\tl{\partial}_{12;s}\to \tl{\partial}_{12}$. Starting from the full ML equation~\eqref{eq:fullMasterEq}, let us apply a variable transformation: 
		\begin{equation}
			\hz_{\CM}:= \frac{\hz_1+\hz_2}{2}, \>\>\> \hr:=\hz_1-\hz_{\CM}=\frac{\hz_1-\hz_2}{2}
			\>\>\> \Longleftrightarrow \>\>\>
			\hz_1 = \hz_{\CM} + \hr, \>\>\> \hz_2 = \hz_{\CM} - \hr. 
		\end{equation}
		This implies the derivative chain rule
		\begin{equation}
			\frac{\partial}{\partial z_1} = 
				\frac{1}{2}\left(\frac{\partial}{\partial z_{\CM}}+\frac{\partial}{\partial r}\right), \>\>\>
			\frac{\partial}{\partial z_2} = 
				\frac{1}{2}\left(\frac{\partial}{\partial z_{\CM}}-\frac{\partial}{\partial r}\right).
		\end{equation}
		We thus obtain 
		\begin{align}
			\frac{\partial P_t(z_{\CM},r)}{\partial t} =& \frac{\sigmaCM^2}{2}\left[\frac{\partial^2}{\partial z_{\CM}^2} + \frac{\partial^2}{\partial r^2}\right]P_t(z_{\CM},r)
			+\sum_{s=\pm 1}\left[J_{t;s}(z_{\CM},r-sL/2) - J_{t;s}(z_{\CM},r)\right] \\
			J_{t;s}(z_{\CM},r) :=& -\frac{s \sigmaCM^2}{2}\delta(r-sL/2)\frac{\partial }{\partial r}P_t(z_{\CM},r).
		\end{align}
		where we have used $\sigmaCM^2=\sigma^2/2$. By introducing the reduced PDF 
		\begin{equation}
			P_t(r) := \int_{-\infty}^\infty dz_{\CM}P_t(z_{\CM},r)
		\end{equation}
		as the result of the marginalisation, we obtain the reduced ML equation~\eqref{eq:masterEq1_confined}. This consistency suggests that the kinetic formulation for the dealer model and the technical calculation therein are very reasonable from various viewpoints. 

\section{Advanced modelling: interaction between traders via the market midprice}
\label{sec:dealerModel_interaction}
	Here we introduce a generalised dealer model by incorporating interaction between traders and then exactly solve the model straightforwardly on the basis of the kinetic approach. 

	\subsection{Dealer model with interaction via the market midprice}
		\begin{figure}
			\centering
			\includegraphics[width=120mm]{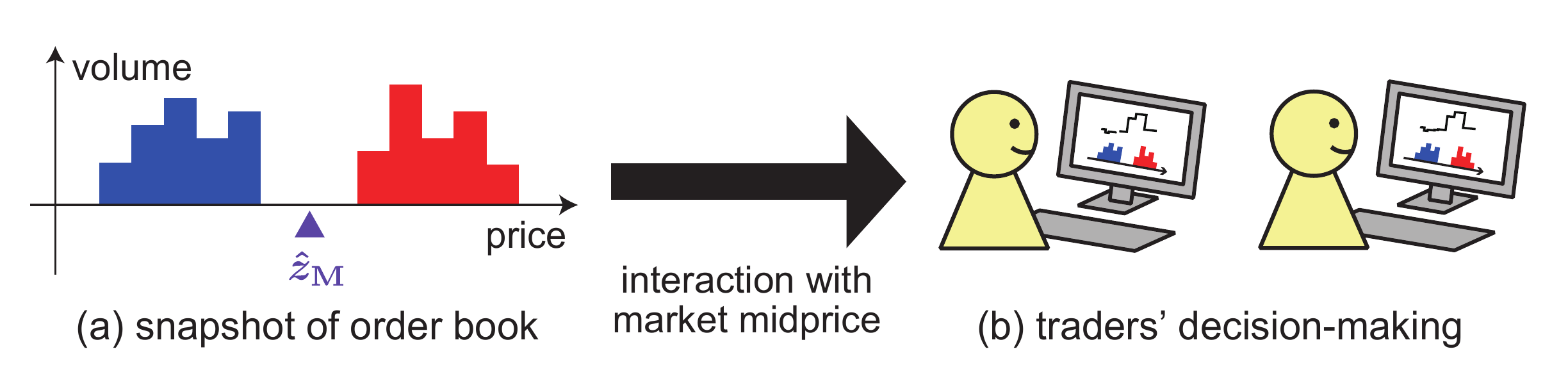}
			\caption{
				Traders make their decisions on the basis of the current order book. One of the key feature variables is the market midprice $\hz_{\rm M}$. As an advanced modelling, interaction between traders and market midprice is incorporated in Sec.~\ref{sec:dealerModel_interaction}.
			}
			\label{fig:interaction_midprice}
		\end{figure}
		We have shown the exact solution to the simple two-body dealer model by the kinetic theory. At the same time, it is realistic to introduce interaction between the traders via the order book. For example, it is a reasonable assumption that traders avoid immediate transactions by submitting orders far from the market midprice $\hz_{\rm M}=\hz_{\CM}=(z_1+z_2)/2$. In the absence of transactions, we thus consider the following generalised dealer model: 
		\begin{subequations}	
			\begin{align}
				\frac{d\hz_1}{dt} &= -\frac{d}{d\hr_1}U\left(\hr_1\right) + \sigma \hxi^{\mrG}_1 \\
				\frac{d\hz_2}{dt} &= -\frac{d}{d\hr_2}U\left(\hr_2\right) + \sigma \hxi^{\mrG}_2
			\end{align}
			with the interaction via the market midprice $U$, the independent white Gaussian noises $\hxi^{\mrG}_i$, and relative prices $\hr_i=\hz_i-\hz_{\rm M}$ from the midprice for $i=1,2$. We assume that the potential is a symmetric function with minimum at $r=0$: 
			\begin{equation}
				U(0) = 0, \>\>\> U(r) = U(-r) \geq U(0) \>\>\> \mbox{for any $r$}.
			\end{equation}
			This potential has the effect to keep the distance between the market midprice and the best bid (ask) price. The market spread tends to be kept wide due to this potential and, thus, immediate transactions are unlikely if the potential strength is strong (see Fig.~\ref{fig:potential_avoiding}). 
			\begin{figure}
				\centering
				\includegraphics[width=135mm]{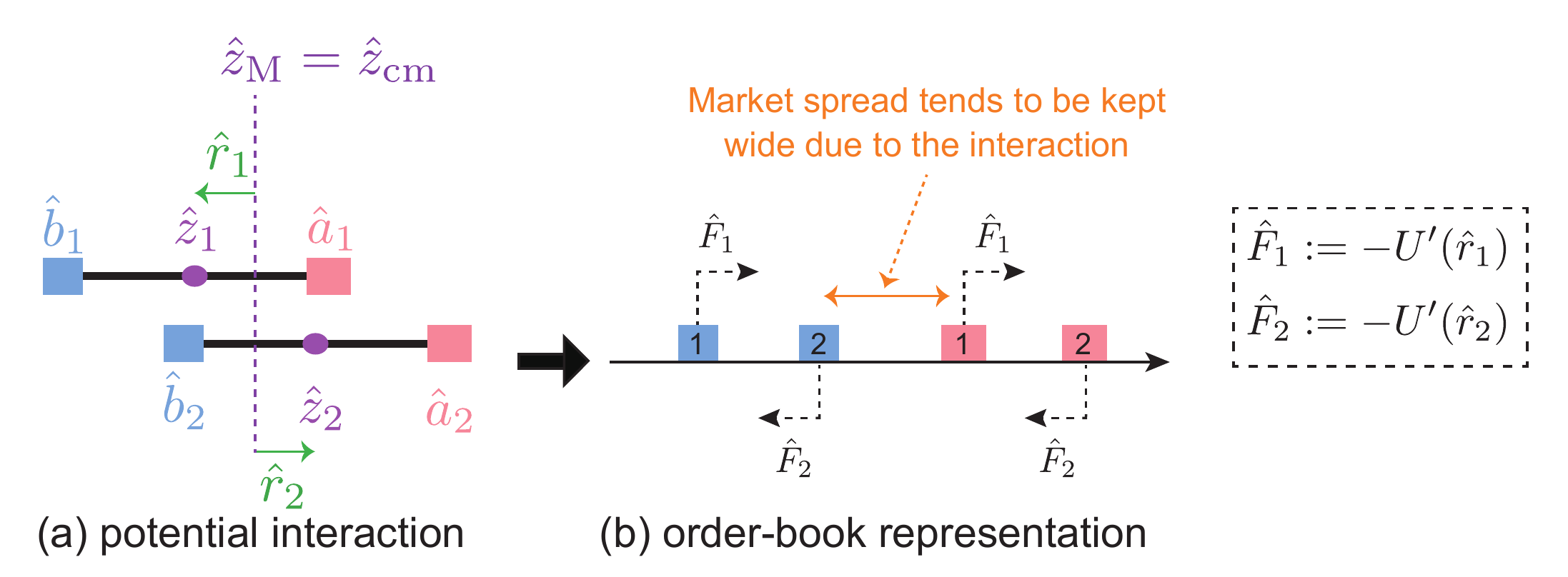}
				\caption{
					Schematic of the potential interaction via the market midprice $\hz_{\rm M}=\hz_{\CM}$. (a) The relative coordinate $\hr_i$ is introduced from the market midprice for $i=1,2$. We can define the ``force" $\hat{F}_i:=-U'(\hr_i)$ which has the effect to keep the market spread wide and thus to prevent immediate transactions.  
				}
				\label{fig:potential_avoiding}
			\end{figure} 
			In the presence of transactions, the jump rule is given by the same rule as the conventional dealer model: 
			\begin{align}
				|\hz_1(t) - \hz_2(t)|=L \>\>\> \Longrightarrow \>\>\> 
				\hz_1(t+dt) = \hz_2(t+dt) = \hz_1(t) - \frac{L}{2}\sgn (\hz_1(t)-\hz_2(t)).
			\end{align}
		\end{subequations}

		We then describe this generalised dealer model in terms of the relative coordinate to the centre of mass $\hr := \hz_1 - \hz_{\CM} = (\hz_1-\hz_2)/2$: 
		\begin{equation}
			r(t+dt) = \begin{cases}
				r(t) + \left(- U'(\hr) + \sigma_{\CM}\heta(t)\right)dt & (|r(t)| < L/2) \\
				0 & (|r(t)|=L/2)
			\end{cases}
			\label{eq:dealer_model_with_interaction_r}
		\end{equation}
		with $\heta(t):= (\hxi^{\mrG}_1-\hxi^{\mrG}_2)/\sqrt{2}$, $U'(r):=dU(r)/dr$, $\sigma_{\CM}:=\sigma/\sqrt{2}$. 
	
	\subsection{ML equation in the presence of $U(r)$}
		We derive the ML equation corresponding to Eq.~\eqref{eq:dealer_model_with_interaction_r}. As a natural extension of the reduced ML Eq.~\eqref{eq:reducedML_probCurrentForm} for $U(r)=0$, we obtain the following reduced ML equation in the presence of the potential: 
		\begin{screen}
			\begin{subequations}
				\label{eq:reducedML_probCurrentForm_w_U(r)}
				\begin{equation}
					\frac{\partial P_t(r)}{\partial t} = -\frac{\partial}{\partial r}\left(j_t^{\rm D}(r)+j_t^{\rm P}(r)\right) + \sum_{s=\pm 1}\left[J_{t;s}(r+sL/2)-J_{t;s}(r)\right], 
					\label{eq:reducedML_probCurrentForm_w_U(r)_1}
				\end{equation}
				where $j_t^{\rm D}(r)$, $j_t^{\rm P}(r)$ and $J_{t;s}(r)$ are the probability currents due to diffusion, potential, and jump, respectively, defined by 
				\begin{equation}
					j_t^{\rm D}(r) := -\frac{\sigma^2_{\CM}}{2}\frac{\partial P_t(r)}{\partial r}, \>\>\> 
					j_t^{\rm P}(r) := -U'(r)P_t(r), \>\>\> 
					J_{t;s}(r) := |j_t^{\rm D}(r)|\delta (r-sL/2) \geq 0,
				\end{equation}
			\end{subequations}
			where we have ignored the minor technical issues on the left and right derivatives.
		\end{screen}

		\subsubsection{Derivation}
			The derivation of Eq.~\eqref{eq:reducedML_probCurrentForm_w_U(r)} is essentially parallel to that of Eq.~\eqref{eq:reducedML_probCurrentForm}. Let us first replace the white Gaussian noise $\heta(t)$ with a coloured Gaussian noise $\heta_{\eps}(t)$, satisfying $\la \heta_{\eps}(t)\ra = 0$ and $\la \heta_{\eps}(t_1)\heta_{\eps}(t_2)\ra = (1/(2\eps))e^{-|t_1-t_2|/\eps}$. The time evolution of an arbitrary function $f(\hr)$ is given by 
			\begin{equation}
				df(\hr) = \begin{cases}
										\displaystyle 
										\frac{df(\hr)}{d\hr}\left(-U'(\hr) + \sigma_{\CM}\heta_{\eps}\right)dt & \mbox{(with a collision: }\htau_{s;i} \not \in [t,t+dt)) \\
										f(\hr-sL/2) - f(\hr) & \mbox{(with a collision: }\htau_{s;i} \in [t,t+dt))
									\end{cases},
			\end{equation}
			where $df(\hr(t)):=f(\hr(t+dt))-f(\hr(t))$ and $\htau_{s;i}$ is the $i$th arrival time of the particle at $\hr(\htau_{s;i})=sL/2$ for $s=\pm 1$. Using the $\delta$ functions, this is equivalent to 
			\begin{align}
				\frac{df(\hr)}{dt} &= \frac{df(\hr)}{d\hr}\left(-U'(\hr) + \sigma_{\CM}\heta_{\eps}\right) + \sum_{s=\pm 1}\sum_{i=1}\left[f(\hr-sL/2)-f(\hr)\right]\delta(t-\htau_{s;i}) \notag \\
				&= \frac{df(\hr)}{dr}\left(-U'(\hr) + \sigma_{\CM}\heta_{\eps}\right) + \sum_{s=\pm 1}\left[f(\hr-sL/2)-f(\hr)\right]\left|\frac{d\hr}{dt}\right|\delta(\hr-sL/2).
			\end{align}
			By considering the sign of the velocity $d\hr/dt$ as $d\hr/dt > 0$ for $\hr=L/2$ and $d\hr/dt < 0$ for $\hr=-L/2$, we obtain $|d\hr/dt|=s(-U'(\hr)+\sigma_{\CM}\heta_{\eps})$ just before the collosion at $\hr=sL/2$. We thus have 
			\begin{equation}
				\frac{df(\hr)}{dt} = \frac{df(\hr)}{d\hr}\left(-U'(\hr) + \sigma_{\CM}\heta_{\eps}\right) + \sum_{s=\pm 1}s\left(-U'(\hr)+\sigma_{\CM}\heta_{\eps}\right)\left[f(\hr-sL/2)-f(\hr)\right]\delta(\hr-sL/2).
			\end{equation} 

			By taking the ensemble average of both hand sides (BHSs), we obtain 
			\begin{equation}
				\left<\frac{df(\hr)}{dt}\right> = \left<\frac{df(\hr)}{d\hr}\left(-U'(\hr) + \sigma_{\CM}\heta_{\eps}\right) + \sum_{s=\pm 1}s\left(-U'(\hr)+\sigma_{\CM}\heta_{\eps}\right)\left[f(\hr-sL/2)-f(\hr)\right]\delta(\hr-sL/2)\right>.
			\end{equation}
			By substituting $f(\hr)=\delta(\hr-r)$, we first obtain the following relations: 
			\begin{equation}
				\left<\frac{df(\hr)}{dt}\right> = \frac{d}{dt}\la f(\hr)\ra = \frac{d}{dt}\int_{-\infty}^\infty dr'f(r')P_t(r') = \frac{\partial }{\partial t}P_t(r),
			\end{equation}
			\begin{equation}
				\left<\frac{df(\hr)}{d\hr}U'(\hr)\right> = \int_{-\infty}^\infty P_t(r')\frac{df(r')}{dr'}U'(r')dr' = -\int_{-\infty}^\infty f(r')\frac{\partial}{\partial r'}\left[U'(r')P_t(r')\right]dr = \frac{\partial}{\partial r}\left[U'(r)P_t(r)\right],
			\end{equation}
			\begin{align}
				\left<U'(\hr)f(\hr-sL/2)\delta(\hr-sL/2) \right>
				&= \int_{-\infty}^\infty dr'P_t(r')U'(r')f(\hr'-sL/2)\delta(r'-sL/2) \notag \\
				&= P_t(sL/2)U'(sL/2)\delta(r) = 0, 
			\end{align}
			and 
			\begin{align}
				\left<U'(\hr)f(\hr)\delta(\hr-sL/2) \right>
				&= \int_{-\infty}^\infty dr'P_t(r')U'(r')f(\hr')\delta(r'-sL/2) \notag \\
				&= P_t(sL/2)U'(sL/2)\delta(r-sL/2) = 0, 
			\end{align}
			where we have used $P_t(sL/2)=0$ (i.e., $P_t(r)$ must have the vanishing probability at the boundary). 
			Using Novikov's theorem and substituting $f(\hr)=\delta(\hr-r)$, in addition, we obtain 
			\begin{equation}
				\lim_{\eps\downarrow 0}\left<\sigma_{\CM}\frac{df(\hr)}{d\hr}\heta_{\eps}\right>
				= \left<\frac{\sigma_{\CM}^2}{2} \frac{df(\hr)}{d\hr^2}\right> = \int_{-\infty}^\infty f(r')\frac{\sigma_{\CM}^2}{2}\frac{\partial^2}{\partial r'^2}P_t(r')dr' = \frac{\sigma_{\CM}^2}{2}\frac{\partial^2}{\partial r^2}P_t(r),
			\end{equation}
			\begin{align}
				\lim_{\eps\downarrow 0}\left<\heta_{\eps}f(\hr-sL/2)\delta(\hr-sL/2)\right>
				&= \left<\frac{\sigma_{\CM}}{2}\frac{\partial}{\partial \hr}f(\hr-sL/2)\delta(\hr-sL/2)\right> \notag \\
				&= \int_{-\infty}^\infty dr' P_t(r')\frac{\sigma_{\CM}}{2}\frac{\partial}{\partial r'}f(r'-sL/2)\delta(r'-sL/2) \notag \\
				&= -\frac{\sigma_{\CM}}{2}\delta(r)\frac{\partial}{\partial r}P_t(sL/2),
			\end{align}
			and
			\begin{align}
				\lim_{\eps\downarrow 0}\left<\heta_{\eps}f(\hr)\delta(\hr-sL/2)\right>
				&= \left<\frac{\sigma_{\CM}}{2}\frac{\partial}{\partial \hr}f(\hr)\delta(\hr-sL/2)\right> \notag \\
				&= \int_{-\infty}^\infty dr' P_t(r')\frac{\sigma_{\CM}}{2}\frac{\partial}{\partial r'}f(r')\delta(r'-sL/2) \notag \\
				&= -\frac{\sigma_{\CM}}{2}\delta(r-sL/2)\frac{\partial}{\partial r}P_t(sL/2).
			\end{align}
			In summary, we have 
			\begin{align}
				\frac{\partial }{\partial t}P_t(r) 
				= \frac{\partial }{\partial r}\left[U'(r) + \frac{\sigma_{\CM}^2}{2}\frac{\partial}{\partial r}\right]P_t(r) + \sum_{s=\pm 1} \left[\delta(r)-\delta(r-sL/2)\right]\left(-s\frac{\sigma_{\CM}^2}{2}\frac{\partial}{\partial r}P_t(sL/2)\right),
			\end{align}
			which is equivalent to Eq.~\eqref{eq:reducedML_probCurrentForm_w_U(r)} by considering that $(\partial/\partial r)P_t(L/2)<0$ and $(\partial/\partial r)P_t(-L/2)>0$.

	\subsection{Exact steady solution for general avoiding potential}
		Here we study the exact solution for the order-book profile for a symmetric general avoiding potential: 
		\begin{screen}
		\begin{subequations}
			\label{eq:order_book_exact_sol_for_general_avoiding_U(r)}
			For a symmetric potential 
			\begin{equation}
				U(r) = U(-r),
			\end{equation}
			we obtain the exact solution
			\begin{equation}
				\phi(r) = \begin{cases}
					\displaystyle
					\frac{1}{Z}\exp\left[-\frac{2U(r)}{\sigma_{\CM}^2}\right]\left\{\mcG(L/2) -\mcG(|r|) \right\} & (|r|\leq L/2) \\
					0 & (|r|>0)
				\end{cases}
			\end{equation}
			with 
			\begin{equation}
				\mcG(r) := \int_0^r dx\exp\left[\frac{2U(x)}{\sigma_{\CM}^2}\right], \>\>\> 
				Z:= \int_{-L/2}^{L/2}dr \exp\left[-\frac{2U(r)}{\sigma^2_{\CM}}\right] \left\{\mcG(L/2) -\mcG(|r|) \right\}.
			\end{equation}					
		\end{subequations}
		\end{screen}		
		
		\subsubsection{Derivation}
			Let us consider the steady solution $\phi(r):=\lim_{t\to \infty}P_t(r)$, which must be symmetric $\phi(r)=\phi(-r)$. For $r \in (0,L/2)$, we obtain 
			\begin{equation}
				\frac{d }{dr}\left[U'(r) + \frac{\sigma_{\CM}^2}{2}\frac{d}{d r}\right]\phi(r) = 0. 
				\label{eq:diff_eq_exact_sol_w_U(r)}
			\end{equation}
			The boundady condition at $r=0$ is given by integrating the BHSs of Eq.~\eqref{eq:reducedML_probCurrentForm_w_U(r)_1} over $(-h,h)$ with $h>0$: 
			\begin{equation}
				0 = \frac{\sigma^2_{\CM}}{2}\left(\frac{\partial}{\partial r}\phi(+h)-\frac{\partial}{\partial r}\phi(-h)\right) - \frac{\sigma^2_{\CM}}{2}\frac{\partial}{\partial r}\phi(L/2) + \frac{\sigma^2_{\CM}}{2}\frac{\partial}{\partial r}\phi(-L/2) + o(h),
			\end{equation}
			which is equivalent to 
			\begin{equation}
				\frac{\partial}{\partial r}\phi(+0) = \frac{\partial}{\partial r}\phi(L/2)
				\label{eq:boundary_condition_r=0_w_U(r)}
			\end{equation}
			for $h\downarrow 0$, by considering the symmetry $(\partial/\partial r)\phi(r)=-(\partial/\partial r)\phi(-r)$. The boundary condition at $r=L/2$ is also given by integrating the BHSs of Eq.~\eqref{eq:reducedML_probCurrentForm_w_U(r)_1} over $(L/2-h,L/2+h)$ with $h>0$: 
			\begin{equation}
				0 = \frac{\sigma^2_{\CM}}{2}\left(\frac{\partial}{\partial r}\phi(L/2+h)-\frac{\partial}{\partial r}\phi(L/2-h)\right) + \frac{\sigma^2_{\CM}}{2}\frac{\partial}{\partial r}\phi(L/2) + o(h). 
				\label{eq:boundary_condition_r=L/2_w_U(r)}
			\end{equation}
			Since $(\partial/\partial r)\phi(L/2+h)=0$, the boundary condition~\eqref{eq:boundary_condition_r=L/2_w_U(r)} always holds for $h\downarrow 0$. 
			Furthermore, the normalisation condition is given by 
			\begin{equation}
				\int_0^{L/2}\phi(r)dr = \frac{1}{2}.
				\label{eq:normalisation_condition_w_U(r)}
			\end{equation}
			By solving Eq.~\eqref{eq:diff_eq_exact_sol_w_U(r)} under the conditions~\eqref{eq:boundary_condition_r=0_w_U(r)} and \eqref{eq:normalisation_condition_w_U(r)}, we obtain the exact solution~\eqref{eq:order_book_exact_sol_for_general_avoiding_U(r)}. 

	\subsection{Exact steady solution for harmonic avoidng potential}
		We next consider the specific case where the avoiding potential is harmonic: 
		\begin{screen}
		For the harmonic avoiding potential
		\begin{equation}
			U(r) = \frac{u^2r^2}{2}, \>\>\> u>0, 
		\end{equation}
		the exact steady solution is given by 
		\begin{equation}
			\phi(r) = \begin{cases}
				\displaystyle
				\frac{1}{Z}e^{-\frac{u^2r^2}{\sigma_{\CM}^2}}\left[\erfi\left(\frac{uL}{2\sigma_{\CM}}\right)-\erfi\left(\frac{u|r|}{\sigma_{\CM}}\right)\right] & (|r|\leq L/2) \\ 
				0 & (|r|>L/2)
			\end{cases}
			\label{eq:order_book_exact_sol_for_harmonic_avoiding_U(r)}
		\end{equation}
		with 
		\begin{equation}
			Z = \frac{1}{2u\sigma_{\CM}\sqrt{\pi}}\left[2\pi\sigma_{\CM}^2\erf\left(\frac{uL}{2\sigma_{\CM}}\right)\erfi\left(\frac{uL}{2\sigma_{\CM}}\right)
			- u^2L^2 {}_2F_2\left({1,1 \atop 3/2,2}\Big|-\frac{u^2L^2}{4\sigma^2_{\CM}}\right)\right].
		\end{equation}			
		\end{screen}
		Here we have introduced the following special functions: 
		\begin{subequations}
			\begin{eqnarray}
				\erf(x) &\equiv& (2/\sqrt{\pi})\int_0^x e^{-t^2}dt\\
				\erfi(x) &\equiv& -i\erf(iz)\\
				{}_2F_2\left({a_1, a_2 \atop b_1,b_2}\Big|z\right) &\equiv& \sum_{n=0}^\infty \frac{(a_1)_n(a_2)_n}{(b_1)_n(b_2)_n}\frac{z^n}{n!}
			\end{eqnarray}
		\end{subequations}
		with the Pochhammer symbol $(a)_n\equiv a(a+1)(a+2)\dots(a+n-1)$ for $n\geq 1$ and $(a)_0=1$.

\section{Numerical confirmation and discussion}
	\label{sec:numerical_confirmation}
	\begin{figure}
		\centering
		\includegraphics[width=150mm]{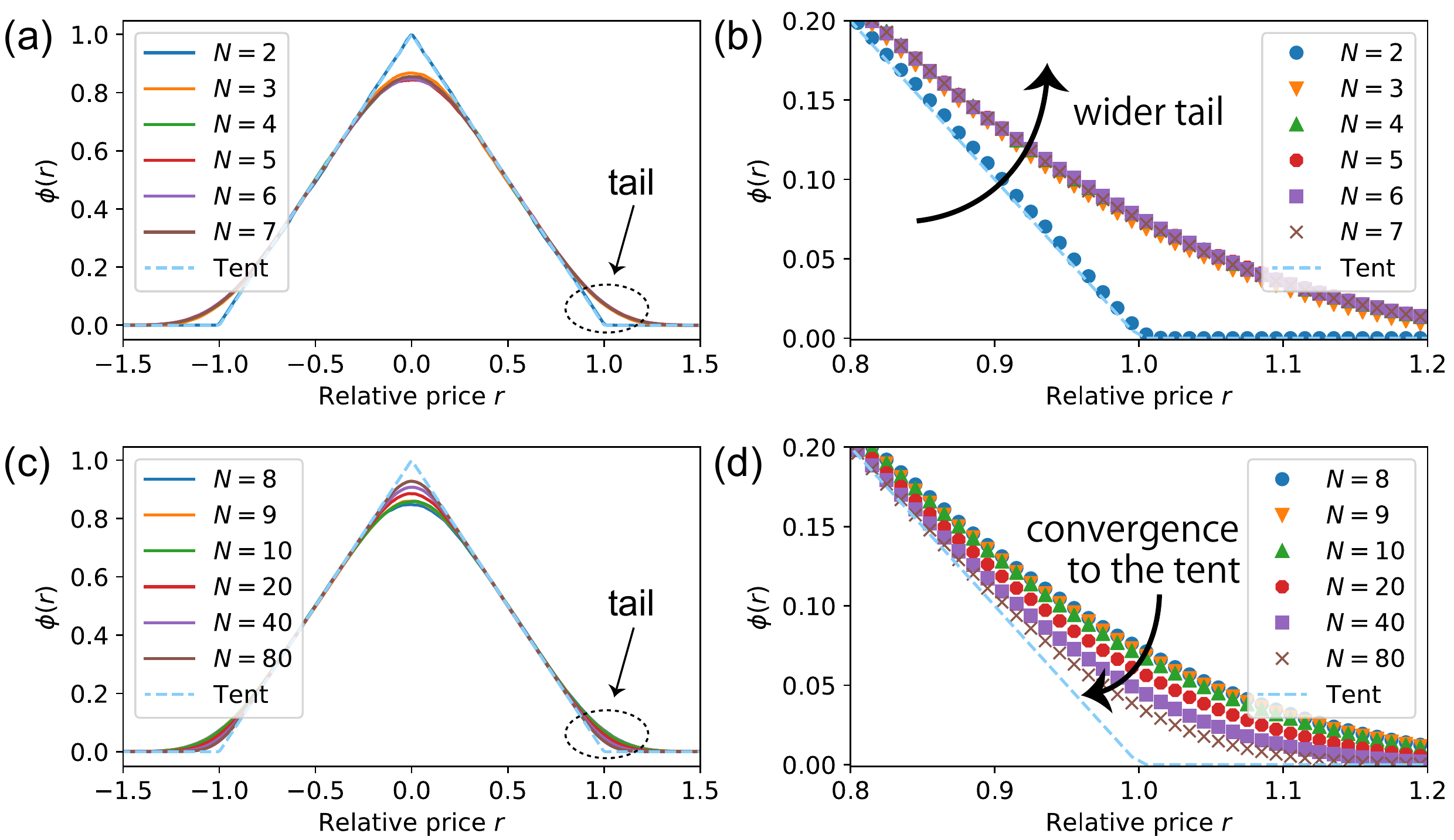}
		\caption{
			Numerical steady PDF $\phi_N(r)$ for the $N$-body dealer model for various $N$, showing the non-monotonic convergence to the tent function~\eqref{eq:tent_N=infty}. (a, b) For $N\in [N,N^*]$, the PDF tail becomes wider with $N^*\simeq 7$, at least numerically. We note that the figure b enlarges the tail near $r=L$. (c, d) The numerical tail seems to exhibit the convergence to the tent function~\eqref{eq:tent_N=infty} for $N \in (N^*,\infty)$. 
		}
		\label{fig:OB_numerical_variousN}
	\end{figure}
	\subsection{Numerical confirmation without avoiding potential}
		\subsubsection{Exact solution for $N=2$}
			Here we numerically confirm the validity of the tent-function formula~\eqref{eq:tent-function_SS} for $\phi(r)$ (see Appendix~\ref{sec:app:numerical_scheme} for the detailed numerical scheme). We have plotted the numerical PDF for $\phi(r)$ as shown in Fig.~\ref{fig:OB_numerical_variousN}a and b, where the tent function is precisely consistent with the numerical result for $N=2$. 

		\subsubsection{Non-monotonic convergence for $N\to \infty$}
			Let us discuss the relationship of the two-body exact solution~\eqref{eq:tent-function_SS} and the numerical solutions for the $N$-body dealer model. According to Refs.~\cite{KanazawaPRL2018,KanazawaPRE2018}, remarkably, the mean-field solution for $N\to \infty$ is also given by the tent function 
			\begin{equation}
				\lim_{N\to\infty}\phi_{N}(r) = \max\left\{0, \frac{L/2-|r|}{L^2/4}\right\},
				\label{eq:tent_N=infty}
			\end{equation}
			where the steady PDF $\phi_N(r):= \la \delta(\hz_i-\hz_{\CM}-r)\ra$ is defined for the $N$-body dealer model, by making assumptions (see Appendix~\ref{sec:app:numerical_scheme} and Refs.~\cite{KanazawaPRL2018,KanazawaPRE2018} for the model assumptions) that 
			\begin{itemize}
				\item all the traders share the same value of their buy-sell spread: $\hL_i:=\ha_i-\hb_i=L=\mbox{const.}$
				\item the dynamics is given by the straightforward generalisation of Eq.~\eqref{eq:dealermodel}; i.e., random walks with ``collisions'' when bid and ask prices coincide with each other. 
			\end{itemize}
			This implies that the exact solution for $N=2$ is equal to the mean-field solution for $N\to \infty$: 
			\begin{equation}
				\phi_{N=2}(r) = \lim_{N\to\infty}\phi_{N}(r).
				\label{eq:equality_btw_N=2AndInfty} 
			\end{equation}
			On the other hand, the solution $\phi_N(r)$ is different from the tent function~\eqref{eq:tent-function_SS} for general $N\neq 2, \infty$: $\phi_N(r):\neq \max\{0,(L/2-|r|)/L^2/4\}$ for $N\neq 2, \infty$. Indeed, the next-leading-order (NLO) mean-field solution for large $N\gg 1$ is given by 
			\begin{equation}
				\phi_{N}(r) = \frac{4\ve}{L^2}\left[\mc{F}\left(\frac{|r|-L/2}{\ve}\right)-2\mc{F}\left(\frac{|r|}{\ve}\right)\right]
				\label{eq:NLO_sol_forN} 
			\end{equation}
			with the thickness of the boundary layer $\ve$ and the tail function $\mc{F}(r)$ defined by 
			\begin{equation}
				\ve:= \frac{L}{2\sqrt{N}}, \>\>\> \mc{F}(r):= \frac{1}{\sqrt{2\pi}}e^{-r^2/2} - \frac{r}{2}{\rm erfc}\left(\frac{r}{\sqrt{2}}\right). 
			\end{equation}
			These relations~\eqref{eq:equality_btw_N=2AndInfty} and \eqref{eq:NLO_sol_forN} suggest that the convergence behaviour of the steady PDF $\phi_N(r)$ is not monotonic in terms of the tail; the tail becomes wider from $N=2$ to $N=N^*$ with some fixed value $N^*>0$ and then it finally converges to the tent function~\eqref{eq:tent-function_SS}. 

			To confirm this picture, we have performed the numerical simulations of the $N$-body dealer model for various $N$ as shown in Fig.~\ref{fig:OB_numerical_variousN} (see Appendix~\ref{sec:app:numerical_scheme} for the detailed numerical scheme). Figures~\ref{fig:OB_numerical_variousN} a and b shows that the tail becomes wider up to $N^*\simeq 7$, numerically. On the other hand, the tail monotonically converges to the tent function~\eqref{eq:tent_N=infty} for $N>N^*$, as suggested by the numerical figures~\ref{fig:OB_numerical_variousN} c and d. This non-monotonic convergence suggests that one of the approximate criteria to apply the mean-field solution~\eqref{eq:NLO_sol_forN} might be to satisfy the condition $N>N^*$ since it is a threshold whereby the solution exhibits qualitatively different behaviours. It might be interesting to investigate the reason behind this non-monotonic convergence numerically and theoretically.

		\subsection{Numerical confirmation under harmonic avoiding potential for $N=2$}
			\begin{figure}
				\centering
				\includegraphics[width=80mm]{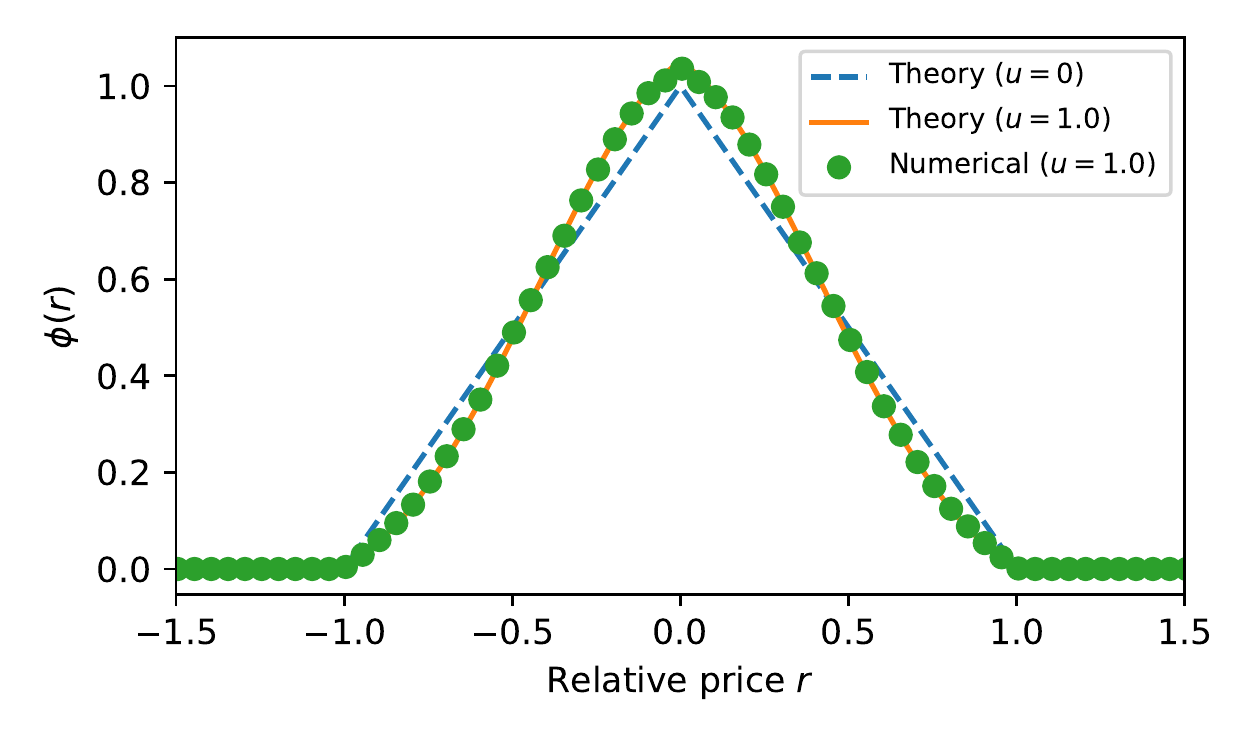}
				\caption{
					Numerical steady PDF $\phi(r)$ for the two-body dealer model under the harmonic avoiding potential $U(r)=u^2r^2/2$ with $u=1$. The numerical PDF $\phi(r)$ excellently fits the theoretical line~\eqref{eq:order_book_exact_sol_for_harmonic_avoiding_U(r)} for $u\neq 0$ but shows discrepancy with the tent function for $u=0$.
				}
				\label{fig:OB_with_potential}
			\end{figure}
			We next confirm our exact solution~\eqref{eq:order_book_exact_sol_for_harmonic_avoiding_U(r)} in the presence of the harmonic avoiding potential for $N=2$ (see Appendix~\ref{sec:app:numerical_scheme} for the detailed numerical scheme). The numerical plot nicely agrees with the theoretical line~\eqref{eq:order_book_exact_sol_for_harmonic_avoiding_U(r)}. In addition, the PDF $\phi(r)$ shrinks around $r=\pm L/2$ in the presence of $U(r)$ and, thus, immediate transactions are unlikely for large $u$.

\section{Conclusion}
	\label{sec:conclusion}
	We have exactly scrutinised the stochastic dealer model by focusing on the specific case $N=2$ from the viewpoint of kinetic theory. We first derive a reduced form of the master-Liouville (ML) equation via two approaches: one is based on Novikov's theorem for coloured noise, and the other is based on a continuous limit from a lattice model. We also examine the physical meaning of the reduced ML equation from the probability current viewpoint, intuitively discerning why the reduced ML equation takes its form as it is. The reduced ML equation is exactly solved to obtain the average-order book profile and the transaction interval. Remarkably, the average transaction interval coincides with that in the previous literature~\cite{YamadaPRE2009}, showing the consistency between the different approaches. We next derive the full ML equation to examine its physical meaning and consistency with the reduced ML equation. To demonstrate the power of this approach, we generalise the dealer model in terms of the interaction between traders via the order book and again exactly solve the generalised model within the kinetic approach. Finally, we provide the numerical simulations to test our exact solution's validity.

  Since the MLEs are derived in this paper, various traditional tools for the master equations will be available for the mathematical analysis of the dealer model. For example, while we have only focused on the steady solution $\phi(r)$, it is possible to consider the time-dependent solutions (since the steady solution corresponds to the ML operator's zero eigenfunction, time-dependent solution corresponds to non-zero eigenfunctions). In addition, it might be interesting to apply the full counting-statistics framework for the MLE to study the complete transaction interval statistics from a different angle. 

  In this paper, we have attempted to thoroughly investigate the mathematical structure of the kinetic theory for financial Brownian motion by focusing on the simplest case of $N=2$. We have shown that various theoretical methods finally produce the same results, which guarantees the mathematical soundness of our approach. While our previous long paper~\cite{KanazawaPRE2018} has meticulously revealed the mean-field mathematical structure of the $N$-body dealer model with $N\gg 1$, this report supplements our previous Letter~\cite{KanazawaPRL2018} from the viewpoint of the exact solution of the simplest case $N=2$, by the detailed description of the ML equations in the complete form. Also, the utility of this mathematical formulation is demonstrated by solving an advanced and realistic dealer model. It would be interesting to observe such a potential interaction from microscopic data analysis directly. In addition, we believe that the thick technical review section would help non-expert readers understand our mathematical theory without hurdles.

\begin{acknowledgement}
	This work was supported by (i) JST, PRESTO Grant Number JPMJPR20M2, Japan, and (ii) JSPS KAKENHI Grant Number 21H01560. We thank Yuki Sato for his reviewing of our manuscript. 
\end{acknowledgement}

\appendix

\section{Brief review on the $\delta$ function}
\label{sec:app:delta_func_review}
	Here we briefly review the $\delta$ function. The $\delta$ function is formally defined by the following relation:
	\begin{equation}
		\delta(x) = \begin{cases}
			0 & (x\neq 0) \\
			\infty & (x=0)
		\end{cases}, \>\>\> 
		\int_{-\infty}^\infty dx f(x)\delta(x) = f(0)
	\end{equation}
	for an arbitrary function $f(x)$. 

	Using this relation, we can derive the variable-transformation formulas. For example, we can derive 
	\begin{equation}
		\delta(ax) = \frac{1}{|a|}\delta(x)
		\label{eq:app:varTrans_delta_linear}
	\end{equation}
	with constant $a\neq 0$. 
	This relation can be derived as follows: for an arbitrary function $f(x)$ and positive $a>0$, we obtain 
	\begin{align}
		\int_{-\infty}^\infty dxf(x)\delta(ax) = \frac{1}{a}\int_{-\infty}^\infty dyf\left(\frac{y}{a}\right)\delta(y) = \frac{1}{a}f(0)
	\end{align}
	by the variable transformation $y:=ax$. The same calculation can be performed for $a<0$ likewise. 

	The variable-transformation formula~\eqref{eq:app:varTrans_delta_linear} can be generalised
	\begin{equation}
		\delta\left(g(x)\right) = \sum_{i} \frac{1}{|g'(x_i)|}\delta(x-x_i)
		\label{eq:app:varTrans_delta_gen}
	\end{equation}
	for an arbitary function $g(x)$, where $x_i$ is the $i$th zero point of $g(x)$ such that $g(x_i)=0$ and $x_i<x_{i+1}$ by assuming $g'(x_i)\neq 0$. This can be derived as follows: for $x$ sufficiently near $x_i$, $g(x)$ can be expanded as 
	\begin{equation}
		g(x) \simeq g(x_i) + g'(x_i) (x-x_i) + o((x-x_i)).  
	\end{equation}
	This means that 
	\begin{equation}
		\delta(g(x)) \simeq \delta\left(g'(x_i)(x-x_i) + o((x-x_i))\right) 
		\simeq \frac{1}{|g'(x_i)|}\delta\left(x-x_i\right)
		\>\>\> \mbox{for $x$ near $x_i$}. 
	\end{equation}
	By considering the contributions for all $x$ near $\{x_i\}_i$, we obtain Eq.~\eqref{eq:app:varTrans_delta_gen}. 

\section{Brief review on the functional Taylor expansion}
\label{sec:app:func_Taylor}
	Here we briefly reivew the functional Taylor expansion. Before the review of the functional Taylor expansion, we first review the Taylor expansion for a $n$-dimensional vector $\bm{x}:=(x_1,\dots,x_{n})$ with a positive integer $n$. For an arbitrary function $f(\bm{x})$, the Taylor expansion implies 
	\begin{equation}
		f(\bm{x}) = \sum_{k=0}^\infty \frac{1}{k!}\left(\sum_{i=1}^n x_i\frac{\partial}{\partial y_i}\right)^k f(\bm{y})\bigg|_{\bm{y}=\bm{0}}. 
		\label{eq:app:functionalTaylor}
	\end{equation}

	The functional Taylor expansion is a generalisation of the relation~\eqref{eq:app:functionalTaylor}. In other words, the functional Taylor expansion implies 
	\begin{equation}
		f[x] = \sum_{k=0}^\infty \frac{1}{k!}\left(\int_{-\infty}^\infty dtx(t)\frac{\delta}{\delta y(t)}\right)^k f[y]\bigg|_{y=0}
	\end{equation}
	with an arbitrary functional $f[x]:=f[\{x(t)\}_t]$ for a function $\{x(t)\}_t$. Here, $\delta/\delta x(t)$ is the functional derivative. 

	The functional derivative $\delta/\delta x(t)$ is related to the $\delta$ function as 
	\begin{equation}
		\frac{\delta x(t)}{\delta x(t')} = \delta(t-t'),
	\end{equation}
	which is a generalisation of the following relation for a finite-dimensional vector $\bm{x}$:
	\begin{equation}
		\frac{\partial x_i}{\partial x_j} = \delta_{i,j}. 
	\end{equation}
	
\section{Numerical scheme for the $N$-body dealer model}\label{sec:app:numerical_scheme}
	\begin{table}[]
		\centering
		\begin{tabular}{llll}
		Variable			& Meaning			                  & Numerical value      	& Dimension \\
		\hline \hline 
		$L$           & Spread constant 							& $2$                  	& Price  \\
		$u^2$         & Strength of the avoiding potential	& $0$ or $1$     	& Time$^{-1}$  \\
		$\sigma^2$    & Variance of the random walks	& $1$                  	& Price$^2$/time \\
		$\Delta t$    & Discrete timestep 						& $10^{-4}/\sqrt{N/2}$ 	& Time \\
		$T_{\rm ini}$ & Time for the initialisation		& $20$                 	& Time \\
		$T_{\rm end}$ & Time for the sampling 				& $10^4$               	& Time \\
		$\Delta r$    & Interval between the bins 		& $10^{-2}$            	& Price \\
		$r_{\max}$    & Maximum value of the bins 		& $3$                  	& Price \\
		$r_{\min}$    & Minimum value of the bins 		& $-3$               		& Price
		\end{tabular}
		\caption{
			Parameter table for the numerical simulations. Here $\Delta t$ depends on the total number of the traders $N$ because the minimum characteristic length $\ve:= L/(2\sqrt{N})$ depends on $N$ likewise. 
		}
		\label{table:numerical_parameters}
	\end{table}
	Here we explain the numerical scheme for the $N$-body dealer model. We also note that the original codes are available as ``Supplementary Code S1.jl'' (without potential $U(r)=0$) and ``Supplementary Code S2.jl'' (with potential $U(r)=u^2r^2/2$), which can be run on the Julia Programming Language. 

	Let us define the bid price, ask price, and midprice of the $i$th trader as $\hb_i$, $\ha_i$, and $\hz_i:=(\hb_i+\ha_i)/2$ for $i \in 1,2,\dots,N$. We assume that the spread is identical to all traders, such that $\ha_i-\hb_i = L = \mbox{const.}$ for all $i = 1,2,\dots, N$. The discrete time is introduced as $t_k = k\Delta t$ with discrete timestep $\Delta t>0$ and integer $k$. We run the simulation during $[-T_{\rm ini},T_{\rm end})$, where the data sample during $[-T_{\rm ini},0)$ is discorded for the initialisation. The dynamical equation is basically given by
	\begin{align}
		\hz_i(t_{k+1}) = \hz_i(t_k) - u^2(\hz_i(t_k)-\hz_{\rm M}(t_k))\Delta t+ \sigma \sqrt{\Delta t}\hxi(t_k), \>\>\> 
		\hz_{\rm M}(t_k) := \frac{1}{2}\left(\hz_1(t_k)+\hz_2(t_k)\right)
	\end{align}
	with the normal random number $\hxi(t_k)$ with the unit variance $\la \hxi(t_k)\hxi(t_{k'})\ra =\delta_{k,k'}$. If there is a transaction, there occurs a jump representing resubmissions: 
	\begin{equation}
		|\hz_i(t_{k}) - \hz_j(t_{k})| \geq L \>\>\> \Longrightarrow \>\>\>
		\hz_i(t_{k+1}) = \hz_j(t_{k+1}) = \frac{\hz_i(t_{k})+\hz_j(t_{k})}{2}. 
	\end{equation}
	Even though it is a rare event for $dt\downarrow 0$, but if there are multiple collisions at the same time, only one pair composed of the highest bid and lowest ask is selected to make a transaction during one discrete timestep. The steady PDF $\phi_N(r)$ is calculated according to the following formula,
	\begin{equation}
		\phi_N(r) := \left<\frac{1}{N}\sum_{i=1}^N \delta(\hz_i-\hz_{\CM}-r)\right>
		\simeq \frac{1}{T_{\rm end}}\sum_{t_k \in [0,T_{\rm end})} \Delta t \left(\frac{1}{N}\sum_{i=1}^N\delta(\hz_i-\hz_{\CM}-r)\right)
		\label{eq:app:PDF_ergodicity}
	\end{equation}
	for sufficiently large $T_{\rm end}$ by assuming the ergodicity. Numerically, we introduce the bins, such as $[r_k,r_{k+1})$ where $r_k:= r_0+k\Delta r$, $r_0 = r_{\min}$, $r_{N_r}=r_{\max}$, and $N_r:=(r_{\max}-r_{\min})/\Delta r$ and then make an approximation on the $\delta$ function as 
	\begin{equation}
		\delta(\hz_i-\hz_{\CM}-r) \simeq 
		\begin{cases}
			1/\Delta r & \mbox{if $r\in [r_k,r_{k+1})$ and $\hz_i-\hz_{\CM} \in [r_k,r_{k+1})$ for some $k \in 0,\dots, N_r-1$} \\
			0 & \mbox{otherwise} 
		\end{cases}.
	\end{equation}
	By applying this binning approximation to Eq.~\eqref{eq:app:PDF_ergodicity}, we numerically evaluate the steady PDF $\phi_N(r)$ under the parameter set summarised in Table~\ref{table:numerical_parameters}.

\end{document}